\documentclass[envcountsame,envcountsect]{llncs}

\usepackage{amsmath,amssymb,stmaryrd,wasysym,url}

\usepackage{amsthm,epsfig,graphicx,booktabs,frameit,color,latexsym,graphics,wrapfig,amssymb,xcolor,colortbl} 

\usepackage[ruled,vlined,algo2e,linesnumbered]{algorithm2e} 

\usepackage{hyperref}
\usepackage[capitalise,nameinlink]{cleveref}

\definecolor{Gray}{gray}{0.85}
\definecolor{darkgreen}{rgb}{0,0.4,0}
\definecolor{darkpurple}{rgb}{0.3,0,0.3}
\colorlet{okcol}{darkgreen}
\colorlet{ercol}{red}
\colorlet{astcol}{blue}

\usepackage{listings}
\newcommand{\savespace}{\vspace{-2mm}}
\newcommand{\saveone}{\vspace{-1mm}}

\crefname{section}{\S{}}{Section}
\crefformat{section}{#2\S{}#1#3}
\Crefformat{section}{Section #2#1#3}

\crefname{lstlisting}{Listing}{Listing}
\crefformat{lstlisting}{#2Listing~#1#3}
\Crefformat{lstlisting}{Listing #2#1#3}

\crefname{definition}{Def.}{Def.}

\crefformat{appendix}{#2\S{}#1#3}
\Crefname{appendix}{Appendix}{Appendix}
\Crefformat{appendix}{Appendix #2#1#3}

\crefname{subsection}{\S\!}{Section}
\Crefname{subsection}{Section}{Section}
\crefformat{subsection}{#2\S{}#1#3}
\Crefformat{subsection}{Section #2#1#3}

\definecolor{bluebell}{rgb}{0.64, 0.64, 0.82}
\newcommand{\bgblue}[1]{{\colorbox{bluebell}{#1}}}

\usepackage{tikz}
\usetikzlibrary{patterns,positioning,backgrounds,calc}
\usepackage{bussproofs,varwidth}
\usepackage{prooftree}
\usepackage{enumitem}

\newcommand{\toolname}{\textsc{CoSl}}
\newcommand{\cslp}{\textsc{CSL-Perm}}
\newcommand{\entname}{\textsc{FrInfer}}

 \theoremstyle{plain}
 \newcommand{\thistheoremname}{}
 \newtheorem*{genericthm}{\thistheoremname}

\renewcommand{\iff}{\Leftrightarrow}
\renewcommand{\vec}[1]{\mathbf{#1}}
\newcommand{\false}{\mathsf{false}}
\newcommand{\true}{\mathsf{true}}
\newcommand{\pre}{\mathsf{pre}}
\newcommand{\post}{\mathsf{post}}

\newcommand{\defeq}{=_{\textrm{\scriptsize{def}}}}

\newcommand{\partialfn}{\rightharpoonup}

\newcommand{\var}{\mathsf{Var}}
\newcommand{\lvar}{\mathsf{LVar}}
\newcommand{\nil}{\mathsf{nil}}
\newcommand{\emp}{\mathsf{emp}}
\newcommand{\psto}[2]{{#1} \mapsto {#2}}
\newcommand{\ppsto}[3]{{#1} \stackrel{{#3}}{\mapsto} {#2}}

\newcommand{\fv}[1]{FV(#1)}

\newcommand{\dom}[1]{\mathrm{dom}\left(#1\right)}

\newcommand{\sem}[1]{\llbracket#1\rrbracket}
\newcommand{\val}{{\sf Val}}
\newcommand{\loc}{{\sf Loc}}
\newcommand{\perm}{{\sf Perm}}

\newcommand{\heaps}{{\sf Heap}}

\newcommand{\sort}[1]{\textsf{#1}}

\newcommand{\Progs}{\sort{Prog}}
\newcommand{\prog}{\ensuremath{\mathsf{P}}}
\newcommand{\FDecs}{\sort{FDec}}
\newcommand{\fdec}{\ensuremath{\mathsf{fd}}}

\newcommand{\Insts}{\sort{Inst}}
\newcommand{\instC}{\ensuremath{\mathsf{inst}}}
\newcommand{\Specs}{\sort{Spec}}
\newcommand{\spec}{\ensuremath{\mathsf{spec}}}
 \newcommand{\skipC}{\code{skip}}

\newcommand{\cassign}[2]{#1\,{:=}\,#2}

\newcommand{\malloc}{\code{malloc()}}
\newcommand{\free}[1]{\code{free}(#1)}
\newcommand{\f}[1]{\code{func(#1)}}

\newcommand{\Comms}{\sort{Comm}}
\newcommand{\comm}{\ensuremath{\mathsf{C}}}
\newcommand{\localsmall}[1]{\code{local}\,#1}
\newcommand{\myif}[3]{\code{if}~#1~\code{then}~#2~\code{else}~#3}
\newcommand{\myatomic}[3]{\code{with}~#1~\code{when}~#2~\code{do}~#3}
\newcommand{\cconcur}[2]{#1||#2}

\newcommand{\ensures}{\mathsf{ensures}}
\newcommand{\requires}{\mathsf{requires}}

\newcommand{\htriplet}[4]{ {#4} \models \mbox{$\htriple{#1}{#2}{#3} $} }

\newcommand{\sllp}{\mathsf{SL}_{\mathsf{LP}}}
\newcommand{\slalp}{\mathsf{SL}_{\mathsf{aLP}}}
\newcommand{\pure}{\mathsf{p}}
\newcommand{\arith}{\mathsf{\phi}}
\newcommand{\Perm}{\mathsf{\Pi}}
\newcommand{\lab}{\mathsf{l}}
\newcommand{\disjoint}{\mathsf{\bot}}
\newcommand{\uweak}{\mathsf{\wcirc}}
\newcommand{\udisjoint}{\mathsf{\circ}}
\newcommand{\heap}{\mathsf{\kappa}}

\newcommand{\labels}{\mathsf{Label}}
\newcommand{\preds}{\mathsf{Pred}}

\newcommand{\wcirc}{\mathrel{\overline{\circ}}}
\newcommand{\wstar}{\circledast}

\newcommand{\entsimpl}[3]{\ensuremath{#1~~{\triangleright}~~#2}}
\newcommand{\minfer}[1]{\bgblue{\ensuremath{#1}}}

\newcommand{\imply}{\ensuremath{\Rightarrow}}
\newcommand{\pto}{{\scriptsize\ensuremath{\mapsto}}}
\newcommand{\seppredF}[2]{\ensuremath{#1(#2)}}
\newcommand{\sepnodeF}[3]{\ensuremath{{#1}{\pto}#3}}
\newcommand{\form}[1]{\ensuremath{#1}}

\newcommand{\ls}[2]{%
\ifx|#2|\then%
\mathsf{ls}%
\else%
\mathsf{ls}(#1,#2)%
\fi}

\newcommand{\tree}[1]{\mathsf{tree}(#1)}

\newcommand{\blue}[1]{\color{blue}{#1}}

\newcommand{\code}[1]{\texttt{#1}}
\newcommand{\assert}[1]{{\blue{\left\{#1\right\}}}}

\newcommand{\hide}[1]{}

\newcommand{\htriple}[3]{\mbox{$\assert{#1}{#2}\assert{#3}$}}
\def\sep{*}

\newcommand{\seppred}[2]{{#1(#2)}}
\def\skip{\code{skip}}

\newcommand{\PSys}{\ensuremath{\Gamma}}
\newcommand{\eqdef}{\defeq}

\newcommand{\exsts}[1]{\exists #1.\;}
\newcommand{\judgetext}{\sort{Eval}}
\newcommand{\unsattext}{\sort{incons}}
\newcommand{\inwardtext}{\sort{inward}}
\newcommand{\inwardstartext}{\sort{inward{\sep}}}

\newcommand{\spost}[2]{\ensuremath{\sort{Post}({#2},{#1}, T)}}
\newcommand{\judge}[2]{\ensuremath{\judgetext(#1, #2)}}
\newcommand{\unsat}[1]{\ensuremath{\unsattext({#1})}}
\newcommand{\wsnorm}[1]{\ensuremath{\sort{ws{-}norm}({#1})}}
\newcommand{\inward}[1]{\ensuremath{\sort{\inwardtext}({#1})}}
\newcommand{\inwardstar}[1]{\ensuremath{\sort{\inwardstartext}({#1})}}
\newcommand{\pureent}[2]{\ensuremath{#1 \Rightarrow_p #2 }}

\newcommand{\conferencepaper}{1} 
\newcommand{\rep}[1]{\ifthenelse{\conferencepaper = 0}{#1}{}}
\newcommand{\repconf}[2]{\ifthenelse{\conferencepaper = 0}{#1}{#2}}

\begin{document}

\title{Compositional Verification with Permissions Regions in Concurrent Separation Logic}
\title{Automated Reasoning over Permissions Regions in Concurrent Separation Logic with Frame Inference}
\title{Frame Inference in Concurrent Separation Logic with Permissions Regions}
\title{Compositional Verification in Concurrent Separation Logic with Permissions Regions}

\author{Quang Loc Le} %
\institute{University College London, United Kingdom} %





\maketitle


\begin{abstract}
  Concurrent separation logic with fractional permissions ({\cslp}) provides a promising reasoning system to verify most complex sequential and
  concurrent fine-grained
  programs. The logic with strong and weak separating conjunctions offers a solid foundation for producing concise
  and precise proofs. 
  However, it lacks automation and compositionality support. This paper addresses this limitation by introducing a compositional verification system for concurrent programs that manipulate
  regions of shared memory. The centre of
  our system is novel logical principles and
  an entailment procedure that can infer the residual heaps in the frame rule for a fragment of {\cslp} with explicit
arithmetical constraints for memory heaps' disjointness. This procedure enables the compositional reasoning for concurrent threads and function calls. We have implemented the proposal in a prototype tool called {\toolname}, tested it with 10 challenging concurrent programs, including those beyond the state-of-the-art, and confirmed the advantage of our approach.
\end{abstract}

\section{Introduction}

This paper introduces a novel compositional verification system for concurrent separation logic with permissions regions
 (\cslp) \cite{Bornat:POPL:2005,Brotherston-etal:20,Le-Hobor:18},
 a promising reasoning system for analysing heap-alternating concurrent programs.
 While
the state-of-the-art {\cslp}, presented in \cite{Brotherston-etal:20}, includes both strong separating conjunction
$\sep$
(to reason about disjointness of heap manipulation) and weak separating conjunction $\wstar$
(to reason on shared resources),
our system takes a step further by providing
 automated and compositional verification.
We demonstrate the effectiveness of our system by applying it to verify 10 challenging fine-grained algorithms.
In the following, we delve into the development of {\cslp} and compositional analysis, as well as our motivations and contributions.

\paragraph{Compositional verification for inter-procedures.}
One of the key challenges in program analysis is {\em local reasoning} about heap-mutating programs.
Separation logic (SL) \cite{OHearn:CSL:2001} provides a well-established framework
for local reasoning.
It includes separating conjunction
$\sep$ to specify the disjointness of heap memory:
$A \sep B$ is a heap that can be split in two separate
parts satisfying $A$ and $B$;
each part, $A$ and $B$, can be analysed independently and locally via
 the following
frame rule.
\[
\AxiomC{$\htriple{P}{C}{Q}$}
\LeftLabel{$(\mathrm{Frame{\sep}-SL})$}
\RightLabel{$(\dagger{:} ~\mathit{Modify(C)} \cap \mathit{FreeVar(F)} = \emptyset)$}
\UnaryInfC{
$\htriple{P \sep F}{C}{Q \sep F}$
}
\DisplayProof
\]
This rule reads: 
The valid Hoare triple $\htriple{P}{C}{Q}$
can be extended to any separated symbolic heap $F$ whenever
the program $C$ does not modify the free variables of $F$.
%
{\em Frame inference} technique further advances local reasoning by inferring
the frame $F$ automatically. By so doing, it enables the compositional verification
and achieves
scalability for inter-procedural analysis \cite{10.1145/3338112}
(i.e., the analysis performance
is proportional to the number of procedures rather than
the number of procedure calls).
Frame inference, the foundation of bi-abduction \cite{Calcagno:POPL:2009},
has established in academic and industrial settings:
The Infer team at Meta implemented it to catch thousands
of bugs in Meta's codebases  \cite{8530713,Le:OOPSLA:2022}.



\paragraph{Compositional verification for concurrent threads.}
There have been some preliminary efforts to transfer compositional verification to concurrency
world using
concurrent separation logic (CSL) \cite{OHearn:CON:2004}.
CSL extends SL
to reason about {\em coarse-grained programs} with concurrent threads (those include
semaphores, conditional critical regions or monitors).
In those programs, each thread owns a separate portion of resources.
 CSL proposed
to utilise the separating conjunction to specify separated ownership of resources:
$P_1 \sep P_2$ is a resource whose ownership can be split into two
parts owned by $P_1$ and $P_2$ separately.
Program analysis can obtain the thread-modular reasoning via the following
inference rule.
\[
\AxiomC{$\htriple{P_1}{C_1}{Q_1}$}
\AxiomC{$\htriple{P_2}{C_2}{Q_2}$}
\LeftLabel{$(\mathrm{Concurrency-CSL})$}
\RightLabel{Non-interference condition}
\BinaryInfC{
$\htriple{P_1 \sep P_2}{C_1 || C_2}{Q_1 \sep Q_2}$
}
\DisplayProof
\]
The rule reads: If $C_1$ and $C_2$ are non-interference concurrent processes (i.e.,
$C_1$ does not modify free variables in $P_2$ and $Q_2$ and vice versa),
then we can analyse these threads independently in the separated ownership
$P_1$ and $P_2$ and combined their analysis results $Q_1$ and $Q_2$
afterwards.
However, the rule above does not work for
{\em fine-grained programs} in which threads interferes each other
 (e.g., a program with two threads sharing read-only variables).




 

To support concurrent fine-grained programs, Bornat {\em et at.} \cite{Bornat:POPL:2005} presented
the first proposal of {\cslp}.
They introduced fractional permission over shared heaps,
e.g., $A^\pi$ is a heap $A$ which owns $\pi$ permission.
If
a thread owns full permission over a resource, it could modify it; otherwise, it
has read-only permission.
Furthermore, the logic specifies shared heaps using
 weak  separating conjunction $\wstar$ (where it is usually written as $\sep$).
 For instance,  $A^{0.5} \wstar A^{0.5}$ means $A$ divides into two parts, each owning
 $0.5$ permissions on $A$.
It also introduces
$(\mathrm{Frame}\wstar)$ and $(\mathrm{Concurrency})$
 by replacing $\sep$ with $\wstar$ in rules $(\mathrm{Frame{\sep}-SL})$ and $(\mathrm{Concurrency-CSL})$
 to reason about shared heaps in concurrent fine-grained programs.
However, this logic 
cannot specify
heaps' disjointness precisely;
for instance, essential entailments such as
$A^{0.5} \sep B^{0.5} \models (A \sep B)^{0.5}$ is valid in CSL but $A^{0.5} \wstar B^{0.5} \models (A \wstar B)^{0.5}$ is false in {\cslp}.
Consequently, it can support neither coarse-grained nor most sequential heap-manipulating programs.

To bridge the gap of {\cslp} \cite{Bornat:POPL:2005},
 Le {\em et} {\em al.} presented  an ad-hoc fix to reason on the disjointness \cite{Le-Hobor:18}.
However, this fix only works for precise formulas \cite{BROOKES2007227}.
Recently,
 Brotherston {\em et} {\em al.} introduced ${\sllp}$,
 a unified separation logic with both strong separating conjunction $\sep$ and
and weak separating conjunction $\wstar$,  that can reason about both
 interfered threads, disjointness of threads and sequential heap-manipulating programs \cite{Brotherston-etal:20}.
${\sllp}$ 
includes heap labels and fractional permissions, e.g., $(\alpha \land A)^\pi$ is a heap $A$ which has label $\alpha$ and owns $\pi$ permission,
and novel logical principles
to retain the heaps' disjointness when switching between $\sep$ and $\wstar$.
However, while this logic demonstrated such a principle
for identical heap formulas, i.e., $(\alpha \wedge A)^{\pi_1} \wstar (\alpha \wedge A)^{\pi_2} \models (\alpha \wedge A)^{\pi_1\oplus\pi_2}$,
it did not show those principles for 
arbitrary formulas. Hence, a proof system based on
${\sllp}$
might not be as strong as the authors had expected.
Moreover,
a study on a frame inference procedure  in ${\sllp}$, which is essential for automation
and compositional reasoning, is still missing.
Since the two conjunctions $\sep$ and $\wstar$ may be alternately nested,
 developing a frame inference procedure
on this logic is non-trivial.
Lastly,  beside of $(\mathrm{Frame}\wstar)$ for $\wstar$,
 ${\sllp}$ includes  another
  frame rule  {$(\mathrm{Frame}\sep)$ for  $\sep$.
Another unanswered question is how the analysis could efficiently decide which frame rule could be applied at a program point.

\noindent {\bf Contributions.}
This paper addresses the aforementioned challenges.
We develop a compositional verification with a frame inference procedure
for $\slalp$,
an analogy of ${\sllp}$
\cite{Brotherston-etal:20}. In our logic, we write
$@_\alpha A^\pi$ for a heap $A$ which has label $\alpha$ and owns $\pi$ permission
and introduce arithmetical operators over heap labels $\alpha$ to keep track of the heaps' disjointness.
These operators are essential to support both frame rules.
Moreover, we make the following primary contributions.




\begin{itemize}
\item We present new distribution principles
 that retain disjointness 
over arbitrary formulas so that
 our system can transform back and forth between the two separating conjunctions.
 \item We propose a frame inference procedure for $\slalp$ that implements the
 distribution principles via inference rules.
 The new principles enable
 our reasoning system
to be more precise than the one in
\cite{Brotherston-etal:20},
 as it could recover $\sep$ from $\wstar$. 
 
 \item We introduce a compositional verification algorithm that supports both thread- and procedure-modular reasoning.
 Our algorithm applies the frame rule with the weak separating conjunction
 and uses the distribution principles to recover strong separating conjunctions.
 By doing so, our system can efficiently support both frame rules.
   %
 We implement a verifier prototype, {\toolname} - the first implemented verification system
 in {\cslp} - to our best knowledge, test it with challenging fine-grained programs, 
 and confirm that {\toolname}
 goes beyond the capability of state-of-the-art logical systems \cite{Bornat:POPL:2005,Brotherston-etal:20,Le-Hobor:18}.
\end{itemize}

\noindent {\bf Outline}
The rest of this paper is organised as follows.
Sect. \ref{sec.mov} illustrates our work through examples.
Sect. \ref{sec:language} shows $\slalp$
 syntax and semantics. 
Sect. \ref{sec:principles} proposes the logical principles
of our frame inference (proofs are in Appendix).
Sect. \ref{sec:ent}
introduces the frame inference procedure with proof rules and a search algorithm.
We present the compositional verification algorithm in
Sect. \ref{sec:analysis}.
Sect. \ref{sec:impl} discusses the implementation
and evaluation.
Sect. \ref{sec.related} shows related work and 
concludes.

\section{Overview and Illustrative Examples} \label{sec.mov}
In this section,
we explain basic concepts, including
 distribution principles, frame inference procedure, and modular verification 
 through
 examples.

\paragraph{Distribution principles.} In practice, sequential code
and concurrent code are often mixed together.
When entering concurrent code from sequential code,
an analysis may split
heap permissions on shared resources
with multiple threads, e.g., using the distribution principle
$@_\alpha A \sep @_\beta B \models @_\alpha A \wstar @_\beta B$ to distribute $\sep$ over $\wstar$ \cite{Brotherston-etal:20}.
Later, when joining the threads and the permissions,
the analysis may want to recover the disjointness for a stronger post-state.
However, $@_\alpha A \wstar @_\beta B \not\models @_\alpha A \sep @_\beta B$
as we have lost information when applying the previous distribution.
In this paper, we propose new logical principles to retain the heaps' disjointness.
For instance, we fix the above distribution as $@_\alpha A \sep @_\beta B = @_\alpha A \wstar @_\beta B \land \minfer{\alpha \disjoint \beta}$
where $\alpha \disjoint \beta$ is a pure/non-heap formula that specifies the heaps' disjointness;
it enables the analysis to recover the disjointness. 

\paragraph{Frame inference.}
We propose a procedure, called {\entname}, to
solve the following frame inference problem:
\saveone\[\saveone
\begin{array}{l}
\text{Given:}~ A \mbox{ and goal } G \\
\text{To find:}~ \mbox{a collection of frame assertions }  F \mbox{ such  that } $$A  \models G \wstar F $$
\end{array}
\]
where $F$ includes the residual heap (of $A$) and arithmetical constraints
that instantiate those logical variables and existentially quantified variables in $G$
with those terms in $A$, and
 $\wstar$ is
 the weak separating conjunction. {\entname} naturally automates
 the frame rule with weak separating conjunction.
 Moreover, {\entname} also implements the distribution principles
 in
 a function, called $\wsnorm{A}$,
 so that $F$ includes disjointness constraints that helps transforming
 $G \wstar F$ to $G \sep F$, when applicable. And so,
 {\entname} can support the automation of the frame rule
 with strong separating conjunction as well.


We illustrate how {\entname} finds frame $F_1$
for the following
entailment problem: 
$@_\alpha \ppsto{x}{y}{1/2} \sep @_\beta \ppsto{y}{\nil}{1/2} \wedge x\neq y \models \exists w. @_\gamma \ppsto{x}{w}{1/2}$.
Here,
$@_\alpha \ppsto{x}{y}{1/2}$, a shorthand for $(@_{\alpha} {x}\mapsto y)^{1/2}$,
says that $x$ points to a heap with $1/2$ permission, named $\alpha$, and its content is $y$; note that, labels $\alpha$, $\beta$
and $\gamma$ are logical variables.

Our entailment procedure {\entname}
 outputs  $\{F_1\}$
with
$F_1 \equiv @_\beta  \ppsto{y}{\nil}{1/2} \wedge x\neq y \wedge \alpha \disjoint \beta \land \minfer{\gamma=\alpha \land w=y}$.
{\entname} infers that if we instantiate the logical variable $\gamma$
with $\alpha$ and the existentially quantified variable $w$ with $y$
(the constraint in the $\minfer{\text{greyed area}}$ shows the instantiation).
 $F_1$ includes the cell pointed to by $y$.
 Furthermore,
$\alpha \disjoint \beta$ specifies the heaps' disjointness
and helps to recover the disjointness whenever $F_1$ is re-combined with the LHS of
the above entailment. 

Similarly, given the entailment, referred to as $e$,
$@_\alpha \ppsto{x}{y}{1/2} \sep @_\beta \ppsto{y}{\nil}{1/2} \wedge x\neq y \models \exists w. @_\gamma \ppsto{x}{w}{1/4}$,
 {\entname}
 returns  $\{F_2\}$ where $F_2 \equiv @_\alpha\ppsto{x}{y}{1/4} \wstar  @_\beta  \ppsto{y}{\nil}{1/2} \wedge x\neq y \wedge \alpha \disjoint \beta \land {\gamma=\alpha \land w=y}$.
The residual heaps include two cells: while the cell pointed to by $x$ remains $1/4 (= 1/2-1/4)$
permission,
the cell pointed to by $y$ is preserved intact.





\paragraph{Compositional verification.}
Our verification system,
{\toolname}, verifies a program bottom-up; it
verifies the callees before the caller and split/join threads before
 the main thread.
It statically executes
a program by calculating symbolical constraints
on values stored in program variables. This execution amounts to
the transformation of the constraints over
program statements whose meanings are given via their abstractions
in the form of Hoare triples.
For instance
 Hoare triple $\htriple{\pre}{C}{\post}$ is the abstraction
 of program $C$ with
pre-condition $\textcolor{blue}{\pre}$ and post-condition $\textcolor{blue}{\post}$,
where $\textcolor{blue}{\pre}$ and $\textcolor{blue}{\post}$ are 
 $\slalp$ formulas.


Suppose $ \{@_\gamma \ppsto{x}{w}{1/4}  \} \code{foo(x)} \{@_\gamma \ppsto{x}{w}{1/4}\}$
is valid,
let us illustrate how {\toolname} verifies the following Hoare triple
at a call site of function \code{foo}.
 $$ \{@_\alpha \ppsto{x}{y}{1/2} \sep @_\beta \ppsto{y}{\nil}{1/2} \wedge x\neq y \} \code{foo(x)} \{@_\alpha \ppsto{x}{y}{1/2} \sep @_\beta \ppsto{y}{\nil}{1/2} \wedge x\neq y\}$$
{\toolname} successfully verifies the triple with the following proof steps.
\begin{enumerate}
\item It invokes {\entname} to find the frame $F_2$ of the entailment $e$ above.
\item It applies weak frame rule to find the post-state.
\saveone\[\saveone
\AxiomC{$\{@_\gamma \ppsto{x}{w}{1/4}  \} \code{foo(x)} \{@_\gamma \ppsto{x}{w}{1/4}\}$}
\LeftLabel{$(\mathrm{Frame}\wstar)$}
\RightLabel{$(\dagger)$}
\UnaryInfC{
$\htriple{@_\gamma \ppsto{x}{w}{1/4} \wstar F_2}{\code{foo(x)}}{@_\gamma \ppsto{x}{w}{1/4} \wstar F_2}$
}
\DisplayProof
\]
\item It applies $\wsnorm{@_\gamma \ppsto{x}{w}{1/4} \wstar F_2}$ to restore the strong separation in the post-state
 and obtain
$@_\alpha \ppsto{x}{y}{1/2} \sep @_\beta \ppsto{y}{\nil}{1/2} \wedge x\neq y
\wedge \alpha \disjoint \beta \land {\gamma=\alpha \land w=y}$.

\item {\toolname} invokes {\entname} again to prove
\saveone\[\saveone
\begin{array}{l}
@_\alpha \ppsto{x}{y}{1/2} \sep @_\beta \ppsto{y}{\nil}{1/2} \wedge x\neq y
\wedge \alpha \disjoint \beta \land {\gamma=\alpha \land w=y} \\
\qquad \qquad \models @_\alpha \ppsto{x}{y}{1/2} \sep @_\beta \ppsto{y}{\nil}{1/2} \wedge x\neq y
\end{array}
\]
{\entname} produces $\{F_3\}$ where $F_3 \equiv \emp \wedge x\neq y \wedge \alpha \disjoint \beta \land {\gamma=\alpha \land w=y}$,
where $\emp$ is the empty heap predicate. As so, it concludes the triple is valid.
\end{enumerate}

 We have tested
 {\toolname} with several fined-grain concurrent programs including those in \cite{Brotherston-etal:20,Le-Hobor:18} and
the one in \cref{fig:ex} which is beyond the capability of the existing systems.
%
Procedure \code{traverse(x, y)}  traverses and processes
 data over  \emph{binary trees} in which \code{traverse(x, y)} accesses two sub-trees
in parallel.
{\blue{$\pre$}} and {\blue{$\post$}}
in the specification
 are defined through
binary trees as:
\saveone\[\saveone
\begin{array}{lcl}
\emp \wedge x=\nil &\Rightarrow& @_{\alpha_0} \seppred{\code{tree}}{x} \\
 \exists d,l,r, \alpha, \beta, \gamma.\ @_\alpha x\mapsto d,l,r \sep @_\beta \tree{l} \sep @_\gamma \tree{r} \wedge \alpha_0 {=} \alpha \udisjoint \beta \udisjoint \gamma &\imply& @_{\alpha_0} \seppred{\code{tree}}{x}
\end{array}
\]
In {\blue{$\pre$}},
$@_\alpha \tree{x}^\pi$ specifies binary trees, rooted by \code{x},
whose heaps are allocated in a memory labelled with $\alpha$.
Line 3 uses the procedure \code{process} to process the
content of pointer $x$ and the tree rooted by $y$,
and on line 4, it recursively traverses the left and right subtrees in parallel.
The specification of \code{process} is:
$\{ @_{\alpha'} \ppsto{x}{d,l,r}{\pi'} \sep @_{\beta'}\tree{y}^{\sigma'} \} \code{process(x, y)} \{@_{\alpha'} \ppsto{x}{d,l,r}{\pi'} \sep @_{\beta'}\tree{y}^{\sigma'} \}$.
(We recap that $@_{\alpha'} \ppsto{x}{d,l,r}{\pi'}$ is a shorthand for $(@_{\alpha'} {x}\mapsto{d,l,r})^{\pi'}$).
Note that, in our system, the heaps' disjointness is captured
explicitly in
the pure part, e.g.,
the heap disjointness constraints
 $\alpha \disjoint \beta$ in {\blue{$\pre$}} and {\blue{$\post$}}.


To verify \code{traverse(x, y)}, {\toolname} starts with the $\pre$ and
statically executes statements in a bottom-up and thread-modular manner.
It generates Hoare triples, where
those intermediate states are shown
at the bottom half of Figure \ref{fig:ex}.
For the conditional statement on line 2,
it generates the post-state {\blue{$s_1$}} for the $\true$ branch
and {\blue{$s_2$}} for the $\false$ branch.
To verify the $\true$ branch,
it derives the triple $\htriple{\blue{s_1}}{\skip}{\blue{s_1}}$.
After that, it invokes {\entname}
to prove $\blue{s_1} \models \blue{\post}$.
Our procedure establishes the validity of this entailment and
infers the frame, the residual heap, as $\emp$
to show that the program does not leak any resource.

For the $\false$ branch,
when encountering the function call of \code{process} on line 3,
{\toolname} uses the \code{process}'s summary to reason about the call.
To derive the state after line 3,
{\toolname} applies the combination of the Consequence
and Frame rules to infer the post-state in the $\minfer{\text{greyed areas}}$
as follows.
\saveone\[\saveone
\AxiomC{${{\blue{s_2}}} \models p \minfer{\wstar ~ F}$}
\LeftLabel{$(\mathrm{Frame}\wstar)$}
\AxiomC{$\htriple{p}{\code{process(x,y)}}{q}$}
\UnaryInfC{$\htriple{p \wstar F}{\code{process(x,y)}}{q \wstar F}$}
\LeftLabel{$(\mathrm{Conseq})$}
\BinaryInfC{
$\htriple{\blue{s_2}}{\code{process(x,y)}}{\minfer{q \wstar F}}$
}
\DisplayProof
\]
This proof rule reads in a bottom-up and left-right manner.
Starting from ${\blue{s_2}}$ in the conclusion, in the left premise, it invokes {\entname}
to infer the frame $F$ that is not used by the precondition ${\blue{p}}$
of the procedure \code{process}. After that, the frame rule in the right-hand side premise is applied,
and $F$ is combined
with postcondition ${\blue{q}}$ before normalising the combination
to obtain ${\blue{s_3}}$.
Concretely, {\toolname} poses the frame inference query ${\blue{s_2}} \models @_{\alpha'} \ppsto{x}{d,l,r}{\pi'}$,
yielding $\{F\}$ with
$F \equiv @_{\alpha_2} \tree{l}^\pi \sep @_{\alpha_3} \tree{r}^\pi \sep @_\beta \tree{y}^\sigma \wedge \alpha \disjoint \beta \wedge x \neq \nil \wedge \alpha = \alpha_1 \udisjoint \alpha_2^\pi \udisjoint \alpha_3^\pi \land e$ where $e$  is an instantiation for logical and existentially quantified variables
and $e \equiv \alpha'= \alpha_1 \wedge \pi'=\pi$.



 \begin{figure}[tb]
\[
\begin{array}{l}
\begin{array}{cl||l}
	 1{:} &  \multicolumn{2}{l}{\code{traverse(x, y) \{}}\\
 & \multicolumn{2}{l}{\code{$ \assert{\pre{:}~@_\alpha  \tree{x}^\pi \sep @_\beta \tree{y}^\sigma \wedge \alpha \disjoint \beta}$}}\\
2{:} &	   \multicolumn{2}{l}{\code{\indent if (x == null) \{ return;  /*$\blue{s_1}{:}~ {@_\alpha \tree{x}^{\pi} \sep @_\beta  \tree{y}^{\sigma}\wedge \alpha \disjoint \beta \wedge x=\nil}$*/ \} }}\\
 &           \multicolumn{2}{l}{// {\blue{s_2}{:}~ @_\alpha \tree{x}^\pi \sep @_\beta \tree{y}^\sigma
\wedge \alpha \disjoint \beta
 \wedge  x \neq \nil}}\\
3{:} &           \multicolumn{2}{l}{\code{\indent process(x, y); //{\blue{$s_3$}}}}\\
4{:} &	    \code{\indent traverse(x->l, y);\ \ }  & \code{\ \ traverse(x->r, y);}\\
5{:} &	   \multicolumn{2}{l}{\code{/*{\blue{$s_4$}}*/\}}}\\
& \multicolumn{2}{l}{\code{$ \assert{\post{:}~@_\alpha \tree{x}^\pi \sep @_\beta \tree{y}^\sigma\wedge \alpha \disjoint \beta}$}}
\end{array} \\
\hline
{\blue{s_3}}{:} @_{\alpha_1} \ppsto{x}{d,l,r}{\pi} {\sep} @_{\alpha_2} \tree{l}^\pi {\sep} @_{\alpha_3} \tree{r}^\pi {\sep} @_\beta \tree{y}^\sigma \wedge
  \alpha \disjoint \beta \wedge x {\neq} \nil \wedge \alpha {=} \alpha_1 \udisjoint \alpha_2^\pi \udisjoint \alpha_3^\pi \\
{\blue{s_4}}{:} @_{\alpha_1} \ppsto{x}{d,l,r}{\pi}  \sep
 \big((@_{\alpha_2} \tree{l}^{\pi} \sep @_\beta \tree{y}^{\sigma_1}) \wstar (@_{\alpha_3} \tree{r}^{\pi} \sep @_\beta \tree{y}^{\sigma_2}) \big)  \\
\qquad \wedge \alpha \disjoint \beta \wedge x \neq \nil \wedge \alpha = \alpha_1 \udisjoint \alpha_2^\pi \udisjoint \alpha_3^\pi   \wedge \sigma=\sigma_1{\oplus}\sigma_2
\end{array}
\]
\savespace \savespace
\caption{Concurrent tree traversal with derived program states $\blue{s_1}$, $\blue{s_2}$, $\blue{s_3}$ and $\blue{s_4}$}
\label{fig:ex}
\vspace{2mm}
\end{figure}

To analyse the concurrent statement on line 4,
{\toolname} applies a combination of the consequence rule, the concurrency rule and the frame rule
to derive ${\blue{s_4}}$.
\saveone\[\saveone\small
\AxiomC{${\blue{s_3}} \models \pre_1 \wstar \pre_2 \minfer{\wstar ~ F}$}
\LeftLabel{$(\mathrm{Concurrency})$}
\AxiomC{$\htriple{{\pre}_1}{C_1}{{\post}_1}$}
\AxiomC{$\htriple{{\pre}_2}{C_2}{{\post}_2}$}
\BinaryInfC{$\htriple{{\pre}_1 \wstar {\pre}_2}{C_1 || C_2}{\post_1 \wstar \post_2 }$}
\LeftLabel{$(\mathrm{Frame}\wstar)$}
\UnaryInfC{$\htriple{{\pre}_1 \wstar {\pre}_2 \wstar F}{C_1 || C_2}{\post_1 \wstar \post_2 \wstar F}$}
\BinaryInfC{
$\htriple{{\blue{s_3}}}{C_1 || C_2}{\minfer{\post_1 \wstar \post_2 \wstar F}}$
}
\DisplayProof
\]
In this derivation, ${C_1} = \code{traverse(x->l, y)}$ and ${C_2} = \code{traverse(x->r, y)}$;
they are recursive calls
in the left thread and the right thread, respectively.
To reason about recursion, {\toolname} uses the summaries of
 callees as hypotheses i.e., the following triples are valid:
\[
\begin{array}{c}
\htriple{\pre_1}{\code{traverse(x->l, y)}}{\post_1} \qquad
 \htriple{\pre_2}{\code{traverse(x->r, y)}}{\post_2}
 \end{array}\]
where $\pre_1 \equiv \post_1 \equiv@_{\gamma_2}  \tree{l}^{\pi_1} \sep @_{\beta_1} \tree{y}^{\sigma_1}
\wedge \gamma_2 \disjoint \beta_1$ 
and $\pre_2 \equiv \post_2 \equiv@_{\gamma_3}  \tree{r}^{\pi_2} \sep @_{\beta_2} \tree{y}^{\sigma_2}
\wedge \gamma_3 \disjoint \beta_2$.

The proof above reads from bottom to up and left to right.
Particularly, in the left-hand side premise, it first generates 
query
${\blue{s_3}} \models \pre_1 \wstar \pre_2$,
where $\pre_1 \wstar \pre_2$ results from the concurrency rule.
{\entname} returns
$\{ @_{\alpha_1} \ppsto{x}{d,l,r}{\pi} \wedge \alpha \disjoint \beta \wedge x \neq \nil \wedge \alpha = \alpha_1 \udisjoint \alpha_2^\pi \udisjoint \alpha_3^\pi \land \sigma = \sigma_1 \oplus \sigma_2 \wedge e\}$ where $e$ is an instantiation for logical and quantified variables
and $e \equiv \gamma_2= \alpha_2 \land
\gamma_3= \alpha_3 \wedge \pi_1 =\pi \wedge \pi_2 =\pi \wedge \beta_1 =\beta
\wedge \beta_2 =\beta $.
After that, by applying the frame rule,
 frame $F$ is conjoined with $\post_1 \wstar \post_2$, the post of the
concurrency rule, to obtain the post of the conclusion.
Afterwards, this post is normalised
to obtain ${\blue{s_4}}$.
We remark that the work presented in \cite{Brotherston-etal:20} does not
maintain the disjointness in frame assertions and
could not derive ${\blue{s_4}}$ on line 5.
Hence, verification of this program is beyond its capability.
%
Finally, {\toolname} derives the triple
$\htriple{\blue{s_4}}{\skip}{\blue{\post}}$ and the query ${\blue{s_4}} \models \post$
yielding $\{\emp\}$. 
As such,
 it concludes
 the entailment is valid and the program complies the spec. 

\section{$\slalp$: CSL with Labels, Permissions,
and Arithmetic}
\label{sec:language}

We assume
$\var$ and $\lvar$ as (countably infinite) sets of program variables
and logical variables, respectively,
and
$\val$ as a set of values.
We presume $\var$ and $\lvar$ are disjoint (i.e., $\var \cap  \lvar = \emptyset$).
$\labels$ is a subset of  $\lvar$ which are variables for heap labels. 
$\nil$ is a special variable in $\var$,
and we overload $\nil$ as a special value in $\val$.
$\loc$ is a set of \emph{heap (memory) locations}
such that $\loc \uplus \{\nil\} \subseteq \val$.

To model shared regions, we define a permission algebra as follow.
We assume a tuple $\langle\perm, \oplus, \otimes, \top\rangle$,
where $\perm$ is a set of permissions, $\top$ is the {\em greatest} permission,
$\oplus$ is a partial binary function on $\perm$,
and $\otimes$ is a total binary function on $\perm$.
$\langle\perm, \oplus\rangle$ forms a partial cancellative communtative semigroup
with the properties divisibility ($\forall \pi \in \perm.~ \exists \pi_1,\pi_2 \in \perm.~ \pi = \pi_1\oplus\pi_2$), total permisson ($\forall \pi \in \perm. ~\pi \oplus \top$ is undefined),
no unit ($\forall \pi_1, \pi_2 \in \perm.~ \pi_1 \neq \pi_1\oplus \pi_2$),
and distribution ($\forall \pi, \pi_1, \pi_2 \in \perm.~(\pi_1\oplus\pi_2)\otimes \pi = (\pi_1\otimes\pi)\oplus (\pi_2\oplus\pi)$).

\label{defn:formula}
We define \emph{formulas} of
 $\slalp$ by the grammar:
\saveone\[\savespace
\begin{array}{rrl@{\hspace{3ex}}l}
\Phi & ::= & \true \mid A \mid @_\alpha A \mid 
\Phi \lor \Phi & \mbox{(formula)} \\
A & ::= & \heap \wedge \pure \mid
 \exists x.~ A   & \mbox{(symbolic heap)} \\
\heap & ::= & \emp \mid x \mapsto \vec{y} \mid P(\vec{x}) \mid \heap * \heap \mid \heap \wstar \heap  \mid \heap^{\Perm} \mid @_\alpha \heap & \mbox{(spatial)} \\
 \pure  & ::= & \arith = \arith \mid \arith \neq \arith \mid \arith  \mid \Perm=\Perm \mid \lab = \lab \mid \lab \disjoint \lab \mid \pure \wedge \pure  & \mbox{(pure/non-heap constraints)} \\
 \arith  & ::= & x \mid k \mid \arith + \arith  & \mbox{(arithmetic)} \\
 \Perm  & ::= & \pi \mid \Perm \oplus \Perm \mid \Perm \otimes \Perm  & \mbox{(permission)} \\
\lab & ::= & \alpha  \mid  \lab ~ \udisjoint~ \lab \mid
\lab ~ \uweak~ \lab  \mid  \lab^{\Perm}  & \mbox{(label)}
\end{array}\]
where $x$ ranges over $\var$, $\pi$ over $\perm$, $P$ over $\preds$, $\alpha$
over $\labels$ and $\vec{x}$, $\vec{y}$ over tuples of variables of length matching the arity of the predicate symbol $P$.

We write $@_\alpha \ppsto{x}{y}{\pi}$ for $(@_\alpha x \mapsto y)^\pi$, and $\arith \neq \arith$ for $\neg(\arith=\arith)$.
Inductive predicates are defined by collection of rules each of which is of the form: $\Phi \Rightarrow P(\vec{x})$.
We use ${\PSys}$ to denote a user-defined system of inductive definitions.
For instance, we define predicate \form{\seppredF{\code{lseg}}{x{,}y}}, that
captures list segments,
 as follows.
\saveone\[\saveone
\begin{array}{lcl}
\emp \land x = y &\Rightarrow&  @_{\alpha} \seppred{\code{lseg}}{x{,}y} \\
 {\exists} q, \beta, \gamma {.} @_\beta \sepnodeF{x}{c_1}{q} \sep
@_\gamma \seppred{\code{lseg}}{q{,}y} \land x \neq y \wedge \alpha = \beta \udisjoint \gamma &\imply&  @_{\alpha} \seppred{\code{lseg}}{x{,}y}
\end{array}
\]

We write $\fv{\Phi}$ to denote the set of free variables in $\Phi$.
Label $\alpha$ and permission $\pi$ are logical variables.
The transformation of a formula with constant permissions
into an equivalent formula with logical variables is straightforward.

We define
a \emph{stack} as a map $s : \var \rightarrow \val$;
a \emph{p-heap}, a \emph{heap-with-permissions}, a finite partial function
 $h : \loc \partialfn_{\mathrm{fin}} \val \times \perm$ that maps locations to value-permission pairs;
and \emph{valuation} a function $\rho$ assigning a single p-heap $\rho(\alpha)$ to each label $\alpha \in \labels$.
The satisfaction relation $s,h,\rho \models \Phi$, where $s$ is a stack, $h$ is a p-heap, $\rho$ is a valuation and $\Phi$ is a formula.

 $\dom{h}$ denotes the \emph{domain} of $h$ i.e., the set of heap addresses on which $h$ is defined.  Two p-heaps $h_1$ and $h_2$ are \emph{disjoint} if $\dom{h_1}\cap\dom{h_2}=\emptyset$; furthermore, they are \emph{compatible} if, for all $\ell \in \dom{h_1} \cap \dom{h_2}$,  $h_1(\ell) = (v,\pi_1)$ and $h_2(\ell) = (v,\pi_2)$ and $\pi_1 \oplus \pi_2$ is defined.
The multiplication $\pi\cdot h$ of a p-heap $h$ by permission $\pi$ is defined by extending $\otimes$ pointwise:
$
(\pi \cdot h)(\ell) = (v, \pi\otimes \pi') \;\;\iff\;\; h(\ell) = (v,\pi')
$. 
\emph{Strong composition} $h_1 \circ h_2$ is standard: 1) if $\dom{h_1}\cap\dom{h_2}=\emptyset$,
$(h_1 \circ h_2)(\ell) = h_i(\ell)$ with  $\ell \in \dom{h_i}\setminus\dom{h_{3-i}}$,
 $i \in \{1,2\}$;
and 2) undefined otherwise.
\emph{Weak composition} $h_1 \wcirc h_2$ is defined:
$(h_1 \wcirc h_2)(\ell)$ is 1) $h_i(\ell)$, if $\ell \in \dom{h_i}\setminus\dom{h_{3-i}}$;
2) $(v,\pi_1 \oplus \pi_2)$, if $h_1(\ell)=(v,\pi_1)$, $h_2(\ell)=(v,\pi_2)$
and $\pi_1 \oplus \pi_2$ is defined;
and 3) undefined otherwise.
We define  $\rho(\lab_1 \udisjoint \lab_1) = \rho(\lab_1) \circ \rho(\lab_2)$,
$\rho(\lab_1 \uweak \lab_1) = \rho(\lab_1) \wcirc \rho(\lab_2)$, and
$\rho(\lab^\pi) = \pi \cdot \rho(\lab)$.

We also assume each predicate symbol $P$ of arity $n$ is given a fixed semantics
$\sem{P} \in (\val^n \times \mathit{\heaps})$, like the one in \cite{Brotherston:POPL:08}, where $\mathit{\heaps}$
is the set of all p-heaps.

The semantics is defined by structural induction on $\Phi$
and is shown in App \ref{csl-semantics}.
 The semantics of \emph{entailment} $A \models B$, where $A$ and $B$ are formulas, is:
For any $s$, $h$, and $\rho$,
if $s,h,\rho \models A$ then $s,h,\rho \models B$.

\section{Logical Principles of {$\slalp$}}\label{sec:principles}
We present logical entailments that show the connection
between the strong $\sep$ and weak $\wstar$ separating
conjunctions. Our system
 uses these entailments
to distribute and transform
formulas back and forth between the two conjunctions.

These entailments set the foundations for
the reasoning about multi-threads.
When encountering concurrent code,
an analysis must split the resources the threads share
i.e., weakening the strong separating conjunction $\sep$ to
the weak separating conjunction $\wstar$.
Then, after exiting the concurrency zone,
the analysis might need to strengthen the resources combined from
these threads i.e., re-introducing the strong conjunction $\sep$
from the weak conjunction $\wstar$.
We propose entailment procedure {\entname} and another procedure
${\wsnorm{A}}$ to compositionally reason about multi-threads.
While {\entname} uses concurrency rules to split resources
for predicate matching, which is essential for an entailment procedure in separation logic,
${\wsnorm{A}}$ strengthens program state $A$,
the output of the procedure {\entname},
to restore strong separating conjunctions $\sep$.

In the following,
we discuss two kinds of distributions:
 $\sep$ and $\wstar$ distribution and permission distribution.
While the former shows how to distribute the
two conjunctions over multiple heap regions,
the latter
presents how to distribute permissions over a single heap region.
Along the way, we describe normalisation operators used through these principled
entailments. 

\paragraph{ $\sep$ and $\wstar$ Distribution.}
Intuitively, using these principled entailments to distribute strong separating conjunctions
over weak ones, our system does not discard the disjointness fact; instead,
it keeps this fact as pure formulas i.e.,
through those pure constraints over heap nominals. In so doing,
our system can deduce stronger program states such that at some point during the analysis,
it can soundly revert the weak separating
conjunction to the original form. 
%
\cref{lemma.sw.distribution}
 formalises the results.
We present the proofs of our results in Appendix.

\begin{lemma}[$\sep$ and $\wstar$ Distribution]\label{lemma.sw.distribution}
  For all $A$, $B$, $C$ and $D$, and nominals $\alpha$, $\beta$,
  $\gamma$ and  $\xi$
\begin{align}
& (@_{\alpha} A \wstar @_{\beta} B) \sep (@_{\gamma} C \wstar @_{\xi} D) \notag \\
& \qquad \qquad \models (@_{\alpha} A \sep @_{\gamma} C) \wstar (@_{\beta} B \sep @_{\xi} D)
    \land (\alpha \uweak \beta) \disjoint (\gamma \uweak \xi)
    \label{swdist} \tag{$\wstar/\sep$} \\
    & (@_{\alpha} A \sep @_{\gamma} C) \wstar (@_{\beta} B \sep @_{\xi} D)
\land (\alpha \uweak \beta) \disjoint (\gamma \uweak \xi) \notag \\
 & \qquad \qquad \models
 (@_{\alpha} A \wstar @_{\beta} B) \sep (@_{\gamma} C \wstar @_{\xi} D)
 \label{wsdist} \tag{$\sep/\wstar$}
\end{align}
\end{lemma}

 \cref{lemma.sw.distribution} shows how our system distributes back
and forth
between the two conjunctions  $\sep$ and $\wstar$. 
Before entering threads,
 \eqref{swdist} is used to divide 
 heaps into
multiple shared regions, each  used by a thread, and
\eqref{wsdist} is for strengthening the combined heaps after exiting the concurrent code.
As shown in this Lemma, the pure constraint $\alpha \disjoint \beta$ in \eqref{swtrans}
enables our systems to strengthen the weak conjunction $\wstar$
to the strong conjunction $\sep$ in \eqref{wstrans}.
By taking $B \equiv C \equiv \emp$, 
we obtain
a weak-strong separation conversion principle as 
\begin{align}
@_{\alpha} A \sep @_{\xi} D & \models @_{\alpha} A \wstar @_{\xi} D \land \alpha \disjoint \xi
  \label{swtrans}
   \tag{$\wstar/\sep$} 
  \\
  @_{\alpha} A \wstar @_{\xi} D \land \alpha \disjoint \xi & \models @_{\alpha} A \sep @_{\xi} D
  \label{wstrans} \tag{$\sep/\wstar$} 
\end{align}

\paragraph{Permission Distribution.}
Our system uses \cref{splitjoin} to split and join permissions.
Unlike 
\cite{Brotherston-etal:20}, our principles support
permissions as
logical variables.

\begin{lemma}[Permission Split and Join]\label{splitjoin}
  For all formula $A$, nominal $\alpha$ and permissions $\pi_1$, $\pi_2$ such that
  $\pi_1 \oplus \pi_2$ is defined:
  \begin{align}
    (@_\alpha A)^{\pi_1 \oplus \pi_2} & \models (@_\alpha A)^{\pi_1} \wstar
    (@_\alpha A)^{\pi_2}
    \label{wsplit} \tag{$\wstar$ Split} \\
      (@_\alpha A)^{\pi_1} \wstar
 (@_\alpha A)^{\pi_2} & \models (@_\alpha A)^{\pi_1 \oplus \pi_2}
 \label{wjoin} \tag{$\wstar$ Join}
\end{align}
\end{lemma}

\paragraph{Procedure {\wsnorm{A}}.}
We now describe how procedure {\wsnorm{A}} works.
To transform the weak separating conjunction
to strong one, {\wsnorm{A}} exhaustively applies the following two scenarios into formula $A$:
\begin{enumerate}
\item It applies  \eqref{wsdist}
  and then \eqref{wjoin} into those sub-formulas of the form $(A^{\pi_1} \sep B) \wstar (A^{\pi_2}  \sep C)$ as: $\wsnorm{(A^{\pi_1} \sep B) \wstar (A^{\pi_2}  \sep C)} = A^{\pi_1\oplus \pi_2} \sep (B \wstar C)$
\item It applies  \eqref{wsdist} into those sub-formulas $(@_\alpha A \wstar @_\beta B) \land \alpha \disjoint \beta$ as:
  $$\wsnorm{(@_\alpha A \wstar @_\beta B) \sep C \land \alpha \disjoint \beta} = (@_\alpha A \sep @_\beta B) \sep C \land \alpha \disjoint \beta$$
  Similarly,  $\wsnorm{(@_\alpha A \wstar @_\beta B) \wstar C \land \alpha \disjoint \beta} = (@_\alpha A \sep @_\beta B) \wstar C \land \alpha \disjoint \beta$.
\end{enumerate}

\paragraph{Procedures {\inwardtext} and {\inwardstartext}.}
Finally, we present procedures that distribute nested permissions
over a heap region with individual weak or strong separating conjunctions.
They are used to normalise a formula after unfolding, i.e.,
an inductive predicate
is substituted by its definition;
while {\inwardtext} is used for the unfolding in the LHS of an entailment,
{\inwardstartext} is for the RHS.
They are based on the following entailments and equivalence.
\begin{align}
(@_\alpha A)^{\pi}  & \equiv  @_{ \alpha^\pi} A^{\pi} \tag{$@^{\pi}$} \label{label.perm}\\
 @_\gamma (@_\alpha A \sep @_\beta B)^{\pi}  \wedge \gamma = \alpha^\pi \udisjoint \beta^\pi & \equiv  @_\alpha A^{\pi} \sep @_\beta B^{\pi} \wedge \gamma = \alpha^\pi \udisjoint \beta^\pi \tag{$\sep^\pi$} \label{sep.pi} \\
   @_\gamma (@_\alpha A \wstar @_\beta B)^{\pi} \wedge \gamma = \alpha^\pi \uweak \beta^\pi  & \models   @_\alpha A^{\pi} \wstar @_\beta B^{\pi} \wedge \gamma = \alpha^\pi \uweak \beta^\pi  \tag{$\wstar^\pi$} \label{wsep.pi}  \\
  (A^{\pi_1})^{\pi_2}  & \models A^{\pi_1 \otimes \pi_2} &
  \label{pdown} \tag{$\otimes$}
\end{align}
In the above entailments and equivalence, {\inwardtext} transforms
the formula in the LHS into the ones in the RHS. As
{\inwardstartext} applies for the RHS of an entailment,
it only uses \ref{sep.pi} to transform
the formula for soundness.
The next section shows how
{\entname} implements these principles via inference rules.

\section{Frame Inference Procedure}\label{sec:ent}
Our entailment
 procedure {\entname} finds a collection of  residual heap assertions $F$
 for a given entailment $A \models G$.
 These assertions are then used to analyse
 programs compositionally. We first describe our {\entname} as a proof
 search algorithm and inference rules.
After that, we
elaborate {\entname} with an example. 

\subsection{{\entname} Entailment Algorithm}


Given an entailment $A \models G$,
the core of {\entname} is the judgement function $\judge{A}{G}$
that computes a collection
of frame assertions $F$ such that
 $F$ is a residual heap left over when discharging
the entailment. 
Note that $F$ also includes some instantiations of
logical and existentially quantified variables in $G$
with some terms
in $A$.
\cref{sound.ent} describes the soundness of
 $\judge{A}{G}$.
 \begin{theorem}[Soundness]\label{sound.ent}
\label{soundness}
For all $A, G$:
\[
	F \in \judge{A}{G} \quad \mbox{implies} \quad A  \models G \wstar F
\]
\end{theorem}

\subsubsection {The $\judgetext$ Function}
We define the function $\judgetext$  using the inference rules shown
in \cref{fig:ent.rules}.  Given an entailment $A \models G$, $\judge{A}{G}$ computes
a disjunctive set of frame assertions as follows.
\begin{itemize}
\item Base case:
      $\judge{A}{\emp \land \true} \eqdef \{A\}$.
\item Inductive case:
For every inference rule in \cref{fig:ent.rules} in the following form:
\[
\AxiomC{$\entsimpl{A_n}{G_n}{L_{F_1}}$}
\AxiomC{...}
\AxiomC{$\entsimpl{A_1}{G_1}{L_{F_n}}$}
\TrinaryInfC{
$\entsimpl{A}{G}{L_{F_1} \cup ... \cup L_{F_n}}$
}
\DisplayProof
\]
$\judge{A}{G}$ takes the conclusion as input and computes the premises as
$$\judge{A}{G} \eqdef \judge{A_1}{G_1} \cup ... \cup \judge{A_n}{G_n}$$
\end{itemize}

\begin{figure}[tb]
\begin{center}
\[
\AxiomC{$\entsimpl{A \wedge \pure}{G \square G'}{L}$}
\LeftLabel{$(\mathrm{Hypothesis})$}
\RightLabel{$\pureent{\pure}{\pure_0}$}
\UnaryInfC{
$\entsimpl{A \wedge \pure}{(G \wedge \pure_0) \square G'}{L} $
}
\DisplayProof
\quad
\AxiomC{$\entsimpl{A'}{G'}{L}$}
\LeftLabel{$(\mathrm{\pi})$}
\RightLabel{$A \models G$}
\UnaryInfC{
$\entsimpl{A^{\pi}  \sep A'}{G^{\pi} * G'}{L} $
}
\DisplayProof
\]
\[
\begin{array}{c}
\AxiomC{
$
 \entsimpl{A ~\square~ \inward{@_\alpha A_{1}^\pi} }{G}{}
$
}
\AxiomC{
...
}
\AxiomC{
$
 \entsimpl{A ~\square~ \inward{@_\alpha A_{n}^\pi} }{G}{}
$
}
\LeftLabel{$(\mathrm{LU})$}
\TrinaryInfC{$
\entsimpl{A ~\square~ @_\alpha \seppred{{P}}{\vec{v}}^\pi}{G}{} $}
\DisplayProof
\\
\mbox{ where } \forall i \in \{1...n\}. A_{i} \Rightarrow  \seppred{{P}}{\vec{v}} \in \PSys
\end{array}
\]
\[
\AxiomC{
$
 \entsimpl{A}{G ~\square~ \inwardstar{@_\alpha A_{i}^\pi}}{}
$
}
\LeftLabel{$(\mathrm{RU})$}
\RightLabel{$
A_{i} \Rightarrow  \seppred{{P}}{\vec{v}} \in \PSys
$}
\UnaryInfC{$\begin{array}{l}
\entsimpl{A}{G ~\square~ @_\alpha \seppred{{P}}{\vec{v}}^\pi}{}
\end{array}$}
\DisplayProof
\]
\[
\AxiomC{$\entsimpl{B \wedge \alpha = \alpha' \land \pi = \pi'}{C}{L} $}
\LeftLabel{$(\mathrm{M})$}
\UnaryInfC{
$ \entsimpl{@_{\alpha}A^{\pi} \wstar B}{@_{\alpha'}A^{\pi'} \wstar C}{L}$
}
\DisplayProof
\quad
\AxiomC{$\entsimpl{@_\beta B \wedge \alpha = \alpha' \land \pi = \pi' \wedge \alpha\disjoint\beta}{C}{L} $}
\LeftLabel{$(\mathrm{M}\sep)$}
\UnaryInfC{
$ \entsimpl{@_{\alpha}A^{\pi} \sep @_\beta B}{@_{\alpha'}A^{\pi'} ~\square~ C}{L}$
}
\DisplayProof
\]
\[
\AxiomC{$\entsimpl{B \land \alpha = \beta \land \pi_1 = \pi_2}{C[z_1/z_2] \land y = t}{L}$}
\LeftLabel{$(\mathrm{M}\mapsto)$}
\UnaryInfC{
$ \entsimpl{@_{\alpha}\ppsto{x}{y,z_1}{\pi_1} \wstar B}{\exsts{z_2} @_{\beta}\ppsto{x}{t, z_2}{\pi_2} \wstar C}{L}$
}
\DisplayProof
\]
\[
\AxiomC{$\entsimpl{@_\beta B \land \alpha = \alpha' \land \pi = \pi' \wedge \alpha\disjoint\beta}{C[z_1/z_2] \land y = t}{L}$}
\LeftLabel{$(\mathrm{M\sep}\mapsto)$}
\UnaryInfC{
$ \entsimpl{@_{\alpha}\ppsto{x}{y,z_1}{\pi} \sep @_\beta B}{\exsts{z_2} @_{\alpha'}\ppsto{x}{t, z_2}{\pi'} ~\square~ C}{L}$
}
\DisplayProof
\]
\hide{For each weak heap ((A * C) +* (B * D)) * F2 in RHS
   1- if (A * C) == (B * D) == G
         then find a strong G * F1 in LHS to split G
        (G^p * F1) ==> (G^p1 +* G^p2) * F1 /\ p = p1+p2

                 F1 |- F2
   -----------------------------------
      (G +* G) * F1 |- (G +* G) * F2
    ----------------------------
       G * F1 |- (G +* G) * F2
}
\[
\begin{array}{c}
\AxiomC{$\entsimpl{@_\xi A_1 \wedge \pi=\pi_1\oplus\pi_2\wedge \beta=\alpha \wedge \gamma=\alpha \land \alpha \disjoint \xi}{B \wstar C \wstar G }{L}$}
\LeftLabel{$(\mathrm{Split}\,\wstar)$}
\UnaryInfC{
$\entsimpl{@_\alpha A^{\pi} \sep @_\xi A_1}{(@_\beta A^{\pi_1} \sep B) \wstar (@_\gamma A^{\pi_2} \sep C) \wstar G}
{L}$
}
\DisplayProof
\end{array}
\]
\hide{   2- find ((A * C) * (B * D)) * F1 in LHS,
   weaken it
        ((A * C) +* (B * D)) * F1 ==> ((A * C) * (B * D)) * F1
capture modal @_((A * C) * (B * D)) in LHS

                         F1 |- F2
  ----------------------------------------------------------
   ((A * C) +* (B * D)) * F1 |- ((A * C) +* (B * D)) * F2
 ----------------------------------------------------------
 ((A * C) * (B * D) ) * F1 |- ((A * C) +* (B * D)) * F2

Note: if we do in RHS and the modal in RHS, it losses
 (later on, we could not recover
  strong heap)
      ((A * C) * (B * D)) * F2 <=== ((A * C) +* (B * D)) * F2
  capture modal @_((A * C) * (B * D)) in  RHS}
%
%
\hide{ 3- find ((A +* B) * (C +* D) ) * F1 in LHS to
      apply distribution +*/*. capture modal
            @_(A +* B) * (C +* D)

                          F1 |- F2
  ----------------------------------------------------------
   ((A * C) +* (B * D)) * F1 |- ((A * C) +* (B * D)) * F2
 ----------------------------------------------------------
 ((A +* B) * (C +* D) ) * F1 |- ((A * C) +* (B * D)) * F2
}
%
%
\[
\AxiomC{$\entsimpl{@_{\alpha_1} A_1 \land \alpha'=\alpha \land \beta'=\beta \land \gamma'=\gamma \land \xi'=\xi \land  (\alpha \uweak \beta) \disjoint (\gamma \uweak \xi) \disjoint \alpha_1}{G}{L} $}
\LeftLabel{$(\wstar/*)$}\label{rdistribution}
\UnaryInfC{
$ \entsimpl{(@_\alpha A {\wstar} @_\beta B) \sep (@_\gamma C {\wstar} @_\xi D) \sep @_{\alpha_1} A_1}{(@_{\alpha'} A {\sep} @_{\beta'} B) \wstar (@_{\gamma'} C {\sep}  @_{\xi'} D)  \wstar G}{L} $
}
\DisplayProof
\]
\hide{For each weak heap ((A * C) +* (B * D)) * F1 in LHS
   1- if (A * C) == (B * D) == l & G
         then find a strong G with the same label l
       s.t. G * F2 in RHS to join G
         (G^p1 +* G^p2) * F1 /\ p = p1+p2 ===> (G^p * F1)

subst into modals as well
                          F1 |- F2
   -----------------------------------------------------
   ((l & G) +* (l & G)) * F1 |- ((l & G) +* (l & G)) * F2
   -----------------------------------------------------
   ((l & G) +* (l & G)) * F1 |- (l & G) * F2
}
\[
\AxiomC{$\entsimpl{(@_\beta B \wstar @_\gamma C) \wstar A_1 \wedge\pi = \pi_1 \oplus \pi_2}{G}{L} $}
\LeftLabel{$(\mathrm{Join}\,\wstar)$}\label{eqn:combine_wstar}
\UnaryInfC{
$ \entsimpl{(@_\alpha A^{\pi_1} \sep @_\beta B) \wstar (@_\alpha A^{\pi_2} \sep @_\gamma C) \wstar A_1 \wedge\pi = \pi_1 \oplus \pi_2}{@_\alpha A^{\pi}  * G}{L}$
}
\DisplayProof
\]
\hide{   2- find ((A * C) * (B * D)) * F2 in RHS
and @_((A * C) * (B * D)) in LHS,
           strengthen
((A * C) +* (B * D)) * F1 ==>  (A * C) * (B * D) * F1.

Jump rule
                         F1 |- F2
  ----------------------------------------------------------
  ((A * C) * (B * D)) * F1 |- ((A * C) * (B * D)) * F2
 ----------------------------------------------------------
 ((A * C) +* (B * D)) * F1 |- ((A * C) * (B * D)) * F2
}
\[
\AxiomC{$\entsimpl{A_1 \wedge  \pure }{G}{L}$}
\LeftLabel{$(\sep/\wstar)$}\label{eqn:jump_star}
\UnaryInfC{
$
\begin{array}{c}
\entsimpl{(@_{\alpha} A * @_{\gamma} C) \wstar ( @_{\beta} B *  @_{\xi} D)  \wstar A_1 \wedge \pure}{(@_\alpha A \wstar @_\beta B) * (@_\gamma C \wstar @_\xi D) \sep G}{} \\
\mbox{ with } \pure = (\alpha \uweak \beta) \disjoint (\gamma \uweak \xi)
\end{array}
$
}
\DisplayProof
\]
%
%
\caption{Entailment proof rules for frame inference ($\square$ is $\sep$ or $\wstar$).}\label{fig:ent.rules}
\end{center}
\vspace{-3mm}
\end{figure}
During analysis,
the set of frame assertions is conjoined with program states to form a disjunctive post-state.
Our system analyses this state in the same way with
the post-state produced from conditional \code{\myif{b}{$\comm_1$}{$\comm_2$}} command;
the verification is successful when all the disjuncts of this post-state are
verified successfully.
Rule $\mathrm{Hypothesis}$ eliminates the pure constraints in the RHS.
To discharge $\pureent{p}{p'}$, we partition it into two separating entailments:
one without labels and another with labels. For the former,
we translate it into the arithmetic satisfiability problem.
 For the latter, we make use of the following entailments:
$\pure \land l \models l$ and $\pure \land (\alpha \uweak \beta) \disjoint (\gamma \uweak \xi) \models \alpha\disjoint\gamma$.

Rules $\mathrm{LU}$ and $\mathrm{RU}$ replace an inductive predicate
by its definition in the LHS and RHS, respectively.
$\mathrm{LU}$ is applied when,
under some substitution, the RHS contains a subformula that
is a part of the definition of the inductive predicate.
It is hoped that, after this unfolding, the subformula is matched and subtracted,
thus progressing the proof search.
 $\mathrm{RU}$ is applied similarly but for the RHS.

Rules $\mathrm{M}$, $\mathrm{M\sep}$,
$\mathrm{M}\mapsto$ and $\mathrm{M\sep}\mapsto$
are used to match subformulas.
While rules $\mathrm{M}$ and $\mathrm{M\sep}$ match identical subformulas,
$\mathrm{M}\mapsto$ and $\mathrm{M\sep}\mapsto$ match heap cells
 pointed to by the same pointer and instantiate its content's
 logical/quantified variables, if applicable. Moreover, $\mathrm{M\sep}$
and $\mathrm{M\sep}\mapsto$ apply \cref{lemma.sw.distribution}
to translate the strong separating conjunction
into the weak one before matching.
The soundness of these rules follows
the soundness of the frame rule.
In particular, when the system does matching, it generates corresponding equality constraints.
The respective equality constraints are put into the LHS of the premise
to instantiate the variables for the matches of existentially quantified and logical variables in the RHS of the conclusion (e.g., variables $z_1$ and $z_2$ in these rules). Note that, due to our normalisation, permission variables are logical; hence, our system puts their equalities in the LHS of the premise.
In contrast, for matching other variables in the RHS of the conclusion, the system has to show that the LHS could establish the matches. As such, it puts the equalities (i.e., y = t) into the RHS of the premise, so our system must prove their validity.
For simplicity, these matching rules include point-to predicates with only two fields.
Extension these rules with other cases is straightforward.

Rules $\mathrm{Split}\,\wstar$ and $\mathrm{\wstar/*}$
split heaps and distribute strong separating conjunction over the weak one i.e.,
 multiple threads can concurrently share the heaps.
Rather than discarding the disjointness,
our system preserves the heap separation in the pure part.
 $\mathrm{Join}\,\wstar$ and $\mathrm{*/\wstar}$ combine shared heaps
and restore strong conjunction from the weak one by exploiting
the disjointness constraints in the pure part.
The soundness of these four rules follows \cref{lemma.sw.distribution} and
\cref{splitjoin}.

\subsubsection{Proof Search Algorithm} Given an entailment $A \models G$,
{\entname} build a forest of proofs.
Initially, it creates a proof tree where the root is $A \models G$. It marks this node as an on-going leaf.
For each on-going leaf, it
attempts to match the entailment with the conclusion of the inference
rules in the following order.

\begin{enumerate}
\item If $G \equiv \emp \wedge \true$, mark this node as done with frame $\{A\}$;
\item For each atomic pure formula of $G$, attempt rule $\mathrm{Hypothesis}$;
\item If $A \equiv B^\pi \land \pure$ and $G \equiv  C^\pi \land \pure'$, apply rule $\mathrm{\pi}$
\item Attempt matching rules $\mathrm{M}$, $\mathrm{M\sep}$, $\mathrm{M}\mapsto$, and $\mathrm{M\sep}\mapsto$, if possible;
\item If $G$ is a conjunction of $\wstar$, match $G$
with the RHS of the conclusion of rule $\mathrm{Split}\,\wstar$, if succeed, apply this rule.
Otherwise, try rule $\wstar/*$;
\item If $G$ is a conjunction of $\sep$, match $A$
with the LHS of the conclusion of rule $\mathrm{Join}\,\wstar$, if succeed, apply this rule.
Otherwise, try rule $\sep/\wstar$;
\item Attempt $\mathrm{LU}$. If it fails, try $\mathrm{RU}$;
\item Lastly, other rules.
\end{enumerate}
For all the rule, except $\mathrm{RU}$, if a rule can be applied, {\entname} extends the tree with new on-going leaves each of which
is one of its premise.
For $\mathrm{RU}$, it clones a new tree before do the above extension.
 If no rule applies, {\entname} searches another proof tree. If it finds a tree with all done leaves, it stops the search
 and collects a disjunctive set of frames at the leaves.
 If there is no remaining tree to search, it returns an unknown exception.

\subsection{Illustrative Example}
We elaborate  {\entname} 
using the following entailment.
\saveone\[\saveone
\begin{array}{l}
@_{\alpha_1} \ppsto{x}{d,l,r}{\pi} {\sep} @_{\alpha_2} \tree{l}^\pi {\sep} @_{\alpha_3} \tree{r}^\pi {\sep} @_\beta \tree{y}^\sigma {\wedge} 
  \alpha {\disjoint} \beta {\wedge} x {\neq} \nil {\wedge} \alpha {=} \alpha_1 \udisjoint \alpha_2 \udisjoint \alpha_3 \\
  \quad \models
  (@_{\gamma_2}  \tree{l}^{\pi_1} {\sep} @_{\beta_1} \tree{y}^{\sigma_1}
{\wedge} \gamma_2 \disjoint \beta_1) {\wstar} (@_{\gamma_3}  \tree{r}^{\pi_2} {\sep} @_{\beta_2} \tree{y}^{\sigma_2}
\wedge \gamma_3 \disjoint \beta_2)
  \end{array}
\]

As the RHS is a conjunction of weak separating operators,
it applies rule $\mathrm{Split}\,\wstar$ to split
the permission of $\tree{y}^\sigma$ in the LHS and then match them
with the ones in the RHS. By so doing, it obtains
the following query and generates
the pure constraints in the \minfer{\text{greyed area}}.
\saveone\[\saveone
\begin{array}{l}
@_{\alpha_1} \ppsto{x}{d,l,r}{\pi} \sep @_{\alpha_2} \tree{l}^\pi \sep @_{\alpha_3} \tree{r}^\pi \wedge 
  \alpha \disjoint \beta \wedge x \neq \nil \wedge \alpha {=} \alpha_1 \udisjoint \alpha_2 \udisjoint \alpha_3 \\
  \quad {\wedge} \minfer{\beta_1 {=} \beta {\wedge} \beta_2 {=} \beta {\wedge} \sigma{=}\sigma_1\oplus \sigma_2}  \models
  (@_{\gamma_2}  \tree{l}^{\pi_1}
{\wedge} \gamma_2 {\disjoint} \beta_1) \wstar (@_{\gamma_3}  \tree{r}^{\pi_2}
{\wedge} \gamma_3 {\disjoint} \beta_2)
  \end{array}
\]
Next, it attempts rule $\wstar/*$ to derive
the following entailment
\saveone\[\saveone
\begin{array}{l}
@_{\alpha_1} \ppsto{x}{d,l,r}{\pi}  \wedge 
  \alpha \disjoint \beta \wedge x \neq \nil \wedge \alpha = \alpha_1 \udisjoint \alpha_2 \udisjoint \alpha_3  \wedge {\beta_1 {=} \beta \wedge \beta_2 {=} \beta \wedge \sigma=\sigma_1\oplus \sigma_2} \\
\qquad  \wedge \minfer{\gamma_2=\alpha_2 \wedge \gamma_3=\alpha_3 \land \alpha_2 \disjoint \alpha_3 }
  \models
  \emp \wedge \gamma_2 \disjoint \beta_1 
\wedge \gamma_3 \disjoint \beta_2
  \end{array}
\]
Finally, it applies rule $\mathrm{Hypothesis}$  to remove all pure constraints in the RHS
\saveone\[\saveone
\begin{array}{l}
@_{\alpha_1} \ppsto{x}{d,l,r}{\pi}  \wedge 
  \alpha \disjoint \beta \wedge x \neq \nil \wedge \alpha = \alpha_1 \udisjoint \alpha_2 \udisjoint \alpha_3  \wedge {\beta_1 {=} \beta \wedge \beta_2 {=} \beta \wedge \sigma=\sigma_1\oplus \sigma_2} \\
  \qquad
  \wedge {\gamma_2=\alpha_2 \wedge \gamma_3=\alpha_3 \land \alpha_2 \disjoint \alpha_3 } \models
  \emp \wedge \true
  \end{array}
\]

Afterwards, {\entname} returns the frame assertions $\{@_{\alpha_1} \ppsto{x}{(d,l,r)}{\pi}  \wedge 
  \alpha \disjoint \beta \wedge x \neq \nil \wedge \alpha = \alpha_1 \udisjoint \alpha_2 \udisjoint \alpha_3 
  \wedge {\beta_1 {=} \beta \wedge \beta_2 {=} \beta \wedge \sigma=\sigma_1\oplus \sigma_2}
  \wedge {\gamma_2{=}\alpha_2 \wedge \gamma_3{=}\alpha_3 \land \alpha_2 \disjoint \alpha_3 }\}$.

\section{Compositional Verification Algorithm} \label{sec:analysis}

\begin{figure}[tb]
\[
\begin{array}{r @{\hspace{4pt}} c @{\hspace{4pt}} ll}
        \Insts \ni \instC  & ::= & 
	\skipC 
        \mid \cassign {[x]} v
        \mid \f{$x,y$}
	& \mbox{Instructions}\\
	&&
        \mid \cassign x \malloc
	\mid \free x 
	  \\
	 \Comms \ni \comm & ::= &   \instC \mid \cassign x E \mid \localsmall x. \comm \mid \cassign x {[y]}
         \mid \comm; \comm  \\
        &&
         \mid \myif{\pure}{\comm}{\comm} \mid \cconcur{\f{$x$}}{\f{$x$}} &  \\
          && \mid \myatomic{r}{B}{C} &
         \mbox{Statements}\\
         \Specs \ni \spec  & ::= & \requires~ A~ \ensures~ A \mid \spec;~ \spec & \mbox{Summary}\\
	 \FDecs \ni \fdec  & ::= & \f{x, y} \spec; \comm \mid \fdec; \fdec & \mbox{Function declarations}\\
	 \Progs \ni \prog & ::= & 
         \fdec & \mbox{Programs}\\
\end{array}
\]
\savespace \savespace
\caption{The programming language.}
\label{fig:language}\vspace{2mm}
\end{figure}

The programming language is shown in \cref{fig:language}.
Instructions include
$\skipC$,  
heap store ($\cassign {[x]} y$), function call (\f{$x,y$}),
memory allocation, and memory disposal.
For simplicity, we consider the function has a single call-by-reference parameter $x$
and a single call-by-value parameter $y$; extending our system to deal with other cases would be straightforward.
Statements comprise \emph{instructions} ($\instC$),  assignment ($\cassign x E$),
local variable declaration ($\localsmall x. \comm$), conditional statement ($\myif{\pure}{\comm}{\comm}$),
heap load ($\cassign x {[y]}$),
sequential composition ($\comm; \comm$),   and concurrent statement ($\cconcur{\f{{x}}}{\f{{x}}}$).
We note that, without loss of generality, we encode the thread body with
a function call to simplify the presentation about thread modularity.
A program $\prog$ includes 
a collection of \emph{function declarations}.
A function declaration is of the form $\f{$x, y$} \spec; \comm$, with
$\spec$ is the function's summary, and
$\comm$ is a \emph{statement} implementing the function body. 
The summary is a collection of the function's pre/post specifications.

\paragraph{Verification Algorithm.} We describe
{\toolname} as a proof search
via predicate transformer semantics.
This algorithm generates post-states using {\entname}
to form intermediate Hoare triples.
{\toolname} carries around a specification table, $T$,
that associates each instruction  $\instC$ with a set of specifications.
Concretely,
the $T(\instC)$ is a sequence of Hoare triples:
$\htriple{p_1}{\instC}{q_1}, \cdots , \htriple{p_n}{\instC}{q_n}$.
The user supplies these specifications of instructions. Note that
the specifications of function calls are obtained from the function declarations.
Presumably, all triples in the specification table are valid,
and {\toolname} verifies the validity of every function declaration:
It verifies its body code against the user-provided specifications.

The centre of our algorithm is the function $\spost{\comm}{p}$. Intuitively, given
a precondition $p$, a command {\comm},
and a specification table $T$, $\spost{\comm}{p}$ computes
a disjunctive set of post-conditions $q$, which results from executing $\comm$
on $p$. If the execution fails, it returns an error.
To formally define the meaning of function $\sort{post}$, we write $\htriplet{p}{\comm}{q}{T}$ to denote that 
if the triples in $T$ are valid Hoare triples,
then the triple $\htriple{p}{\comm}{q}$ is also valid.

 \begin{theorem}[Soundness]\label{sound.post}
\label{soundness}
For all $p, \comm, T, q$:
\savespace\[\savespace
	q \in \spost{\comm}{p} \quad \mbox{implies} \quad \htriplet{p}{\comm}{q}{T}
\]
\end{theorem}
 
Given a function declaration $f(x,y) = \comm$ and a
pre/post specification $(\pre, \post)$, {\toolname} invokes $\spost{\comm}{\pre}$
to compute a set of poststates $q$.
After that, by the consequence rule, it checks for all such $q \in \spost{\comm}{\pre}$,
whether $q \models \post$ holds. If this is the case, it concludes that
memory accesses in $\comm$
are safe and that implementation complies with the specification $(\pre, \post)$.
Otherwise, for any $q$ does not entail $\post$, the verification fails.
We define the function $\sort{Post}$ using the inductive rules shown
in \cref{fig:sem}.
These rules are an analogue of
proof rules in concurrency separation logic \cite{OHearn:CON:2004,Brookes:CON:2004,Viktor:TCS:2011} and 
designed to maintain the soundness of the verification described in
\cref{sound.post}.
In rule $\spost{\instC}{p}$, $y$ is the pass-by-value parameter of
a function call. By substituting such a parameter with a fresh variable in the post-condition,
our system forgets its value in the outputs.

\begin{figure}[tbh]
\[
\begin{array}{@{} r @{\hspace{1pt}} c @{\hspace{1pt}} l @{}}
  \spost{\comm}{p} &\eqdef & \{\} \qquad \mbox{ if } $\unsat{p}$ \mbox{ is true}
  \\[1ex]
   \spost{\skipC}{p} &\eqdef & \{p\}
  \\[1ex]
 \spost{\cassign x E}{p} &\eqdef &  \{\exsts{x'} p[x'/x] \wedge x=E[x'/x]\}
 \\[1ex]
\spost{\mathit{\localsmall x. \comm}}{p}
	& \eqdef
	& 
  \{ \exsts{x'} q \mid q \in \spost{\comm[x'/x]}{p} \}
  \\[1ex]
\spost{\cassign x {[y]}}{p}
	& \eqdef
& \{ \sort{ws{-}norm}(\ppsto{y}{y'}{\pi} \wstar F[x'/x] \wedge x=y')\\
  ~~&&\mid~ F \in \entname(p,\ppsto{y}{y'}{\pi})
\}
\\[1ex]
\spost{\cassign x \malloc}{p}
	& \eqdef
& \{ p \land x = \nil; p \sep \exsts L @_\beta \ppsto{x}{L}{1}
\}
\\[1ex]
%
%
	\spost{\myif{b}{\comm_1}{\comm_2}}{p}
	& \eqdef
	&\spost{\comm_1}{p \wedge b}  ~\cup~ \spost{\comm_2}{p \wedge \neg b}
\\[1ex]
\spost{\comm_1;\comm_2}{p}{q}
	& \eqdef
	& \bigcup \{ \spost{\comm_2}{q} \mid q \in \spost{\comm_1}{p} \}
\\[1ex]
  \spost{\instC}{p}
	 &\eqdef&
	 \{ \sort{ws{-}norm}(b[y'/y] \wstar F)
  \mid~ \htriple a \instC b \in T   \\
 && \wedge ~ F \in \entname(p,a)
\}
\\[1ex]
\spost{\cconcur{f(x_1)}{g(x_2)}}{p}
	 &\eqdef&
 \{ \sort{ws{-}norm}((b_1 \wstar b_2) \wstar F) \mid \htriple {a_1}{f(x_1)}{b_1} \in T \\
   && \wedge~
    \htriple{a_2}{g(x_2)}{b_2} \in T  \wedge
    F \in \entname(p,a_1 \wstar a_2)
    \}
   
    \end{array} 
\]
\savespace \savespace
\caption{Predicate transformers for program statements ($x', y', \beta$ are fresh variables).}
\label{fig:sem} \vspace{2mm}
\end{figure}

Function {$\unsat{A}$} checks the inconsistency of $A$
by computing an arithmetical formula $\pure$, which is an over-approximate of $A$.
Given a stack $s$,
 $s' \supset s$ is defined as for all $v \in \dom{s}$, $v \in \dom{s'}$ and $s'(v) = s(v)$.
$\pure$ is an over-approximate of $A$
if there exists $s, h, \rho$ such that $s, h, \rho \models A$, then there exists $s' \supset s. ~s', h, \rho \models \pure$.
Suppose $A \equiv \exists w_1,..,w_n.~ \heap \land \arith \land \lab$,
then $\pure \equiv \exists w_1,..,w_n.~\pure'$, where $\pure'$ is computed as
(we consider $p$ and $p'$ as conjunctive sets of subformulas)
\begin{itemize}
\item If $\ppsto{x}{\_}{\pi} \in \heap$, then $x\neq\nil \in \pure'$.
\item If $\ppsto{x}{\_}{\pi_1} \sep \ppsto{y}{\_}{\pi_2}  \in \heap$, then $x \neq y \in \pure'$.
\item If $@_\alpha \ppsto{x}{\_}{\pi_1} \wstar @_\beta \ppsto{y}{\_}{\pi_2}  \in \heap$ and either $\alpha \disjoint \beta \in \lab$ or $\alpha \udisjoint \beta \in \lab$,
  then $x \neq y \in \pure'$.
\item  
For permission algebra, we make use of Boyland fractional permission model \cite{Boyland:SAS:2003}
where $\perm = \{\pi \mid \pi \in \mathbb{Q} \land 0 < \pi \leq 1\}$,
$\top = 1$, $\oplus$ is $+$ (only defined if the result does not exceed $1$),
and $\otimes$ is $\times$. As so, we replace all $\oplus$ (resp. $\otimes$) in $\arith$
by $+$ (resp. $\times$) to obtain $\arith'$
and $\arith'$ is a conjunct of $\pure'$.
Furthermore, if any permission
$\pi_1+...+\pi_n \in \arith'$, then $0< \pi_1+...+\pi_n \leq 1 \in \pure'$.
\end{itemize}

The summary of instructions are defined as follows
(where $\beta$ is a fresh label).
\saveone\[\saveone
\begin{array}{
c
  }
  
  \begin{array}{ccc}
     \htriple{
	\emp
}{
	\skipC
}{
	 \emp
}
  
     &\quad
    \htriple{
	@_\alpha \ppsto{x}{\_}{1}
}{
	\cassign {[x]} v
}{
	 @_{\beta} \ppsto{x}{v}{1}
}
    &\quad
    \htriple{
	@_\alpha \ppsto{x}{\_}{1}
}{
	\free x
}{
	\emp
}
%
%
%
%
\end{array}
\end{array}
\]
Note that in rule of ${\cassign x \malloc}$, the post-condition describes two cases: ${x = \nil}$ when the system
it out of memory or allocation error and allocation succeeds with a location assigned to \code{x}.
The two last rules, memory write and memory de-allocation, require
full permission in the pre-conditions.
Further note that for those rules using entailment procedure {\entname},
{\toolname} also has to check the side condition of the frame rule.
If either the check
fails with an unknown exception, function ${\sort{Post}}$ returns an error.
After combining the frame $F$ with the post-condition, we further note that these rules call
procedure $\sort{ws{-}norm}$ to transform \underline{\bf w}eak separation $\wstar$ into \underline{\bf s}trong separation
$\sep$, when applicable.



\section{Implementation and Evaluation}\label{sec:impl}

We have implemented {\entname} and {\toolname} using OCaml and Z3 \cite{TACAS08:Moura} as a back-end SMT solver for the arithmetic.
Our implementation supports \code{fork(proc, $\vec{v}$)}/\code{join(id)}.
A fork takes a procedure name \code{proc} and parameters $\vec{v}$ as inputs,
creates a new thread executing the procedure \code{proc}, and returns an identifier of
the newly-created thread. \code{join(id)} waits for the thread \code{id} to finish its execution. For \code{fork}, {\toolname} uses {\entname} to find the frames between
the current program state and
the pre-condition of \code{proc}. For \code{join}, it conjoins using $\wstar$ (and
then normalize)
the current program state with the post-condition of \code{proc}.

Furthermore, we enhance the inference rules in
 {\entname}
by maintaining a data structure that captures
equivalence classes of variables in every pure formula. For example, with
the entailment $@_\alpha \psto{x}{\nil} \land x=y \land y = z \models @_\beta \psto{z}{\nil}$,
it embeds the pure formula in the LHS with
an equivalence class as a set $\{x,y,z\}$.
To apply rule
$\mathrm{M}\mapsto$, 
{\entname} checks whether
the two points-to predicates of the two side pointed to by the same variable.
The equivalence class of the LHS contains three elements,
so it re-arranges the head of the points-to predicate and
tries the following three entailments:
 $@_\alpha \psto{x}{\nil} \land x=y \land y = z \models  @_\beta \psto{z}{\nil}$,
 $@_\alpha \psto{y}{\nil} \land x=y \land y = z \models  @_\beta \psto{z}{\nil}$, and
 $@_\alpha \psto{z}{\nil} \land x=y \land y = z \models  @_\beta \psto{z}{\nil}$.
Since the two points-to predicates in the last entailment match,
the original entailment $@_\alpha \psto{x}{\nil} \land x=y \land y = z \models @_\beta \psto{z}{\nil}$ holds.
%

 We have tested
 {\toolname} with 10 challenging fine-grained concurrent programs. They consist of
 fork/join benchmarks taken from  \cite{Le:ATVA:2013}
  and programs
with complex heap sharing within list segments
or binary trees from \cite{Brotherston-etal:20,Le-Hobor:18}.
We augment parameters of each procedure in the latter set of benchmarks
to ensure the necessity of frame inference during the verification
shared threads.
Furthermore, to increase the complexity of the heap-sharing,
we extend the later set in two ways.
Firstly, we rewrite the program and its specification
using data structures with pure properties like size of lists/trees,
boundaries, and maximum. For instance, instead of traversing a binary tree,
we rewrite the program to count the number of cells allocated in the tree.
Secondly, to test how well our system can handle nested separations,
in some procedures,
we add additional parameters which points to weak or strong separating heaps
with the original ones.
The later set of benchmarks is beyond the capability of \cite{Boyland:SAS:2003,Brotherston-etal:20,Le:ATVA:2013,Le-Hobor:18}.

\begin{table}[t]
\caption{Compositional verification of fine-grained concurrent programs.}
\savespace \savespace \savespace
\label{tbl:expr:veri}
\begin{center}
\begin{tabular}[t]{|c |c |c |c |c | c| }
\hline
 Program/Data Structure(pure properties) & \#LoC & \#LoS & \#procedure & \#query  \\
\hline
\rowcolor{Gray} fibonacci & 38 & 8 
 & 2 
 & 8
  \\
  quicksort using singly linked list(size, min, max) & 99 & 24
 & 3 
 & 17
  \\
 \rowcolor{Gray} mergesort using singly linked list(size, min, max) & 128 & 22
 & 6 
 & 32
  \\
 Singly linked list & 29 & 8 
 & 2 
 & 8
  \\
\rowcolor{Gray} Singly linked list (size) & 45 & 11 
 & 3 
 & 13
  \\
 Skip list (size)
 & 63 & 14
 & 3 
 &   18  \\
\rowcolor{Gray} Doubly linked list & 29 & 8 
 & 2 
 & 8 \\
 Binary trees & 44 & 13 
& 3 
  & 17  \\
\rowcolor{Gray} Binary trees (size) & 57 & 13 
& 4 
& 23    \\
Heap trees (size, maxelem) & 57 & 13 
& 4 
&   23  \\
\hline
\end{tabular}
\end{center}
\end{table}

The experimental results are shown in Table \ref{tbl:expr:veri}.
Each line on this table reports the experiments on one fork/join program or one kind of data structures with
 procedures, number of lines of code (\#LoC),
 number of lines of specification (\#LoS, including the definitions of inductive
 predicates), number of procedure (\#procedure), and
 number frame inference queries (\#query). 
%
Each program includes two procedures
\code{check(x)} and \code{traverse(x,y)}.
\code{check(x)} reads data in a list (or tree) in parallel.
For a list, \code{traverse(x,y)} reads data in a thread and recursively traverses
the tail of the list \code{x} in another thread. For a binary tree,
\code{traverse(x,y)} reads data in both threads and recursively traverses
its two sub-trees in parallel.
Each program with size property includes additional procedure
\code{count(x)}.
For a list, \code{count(x)} reads data in a thread and recursively counts
the tail of the list \code{x} in another thread.
For a skip list, \code{count(x)} counts cells in each level by one thread.
For a binary tree, \code{count(x)} counts cells in the left and right of the tree \code{x}
in parallel.
On a machine with 4 cores and 8GB RAM: {\toolname} verified every program successfully
within {\em one} minute.
The experiments confirm that the frame inference procedure {\entname}
enables {\toolname} to support both procedure-modular and thread-modular.
Furthermore, {\toolname} can produce proofs compositionally and
is potential for scalability.

\section{Related Work and Conclusion} \label{sec.related}
Our work innovatively extends the reasoning system over
${\sllp}$
\cite{Brotherston-etal:20}.
It surpasses \cite{Brotherston-etal:20} by introducing modular reasoning
using a frame inference procedure and an implemented compositional
analysis algorithm. 
Furthermore, our principles on distribution among
sharing conjunction $\wstar$ and separating conjunction $\sep$ are more precise than
the ones in \cite{Brotherston-etal:20}. These principles
help our system to generate stronger post-states during
the verification and thus make {\toolname} more powerful.

Many revisions of CSL have been
proposed to analyse interference and shared resources in
concurrent fine-grained programs.
Their representatives include {\cslp} - the family of
CSL with permissions regions \cite{Bornat:POPL:2005,Brotherston-etal:20,Le-Hobor:18},
CSL$+$Rely/Guarantee \cite{Vafeiadis:CONCUR:2007},
CaReSL \cite{CaReSL:ICFP:2013}, 
iCAP \cite{ICAP:ESOP:2014}, and
Views \cite{Dinsdale:POPL:2013}, and ``protocol-based” notions of disjointness
 in program logic such as
FCSL \cite{Nanevski2014CommunicatingST} and Iris \cite{10.1145/2676726.2676980}.
 {\cslp} are amenable to automation as they
 are simpler and more uniform than protocol-based logics.
Furthermore, while a few could support automation
(for instance, SmallfootRG \cite{SmallfootRG:SAS:2007} for CSL$+$Rely/Guarantee,
Caper \cite{Dinsdale:ESOP:2017} for CAP family, and
Starling \cite{Windsor:CAV:2017} for Views),
 only {\cslp} has demonstrated the compositionality with a frame inference procedure.
 
Our modular verification system for concurrent programs closely relates to
Smallfoot \cite{Smallfoot:FMCO:2005}, jStar \cite{Distefano:2008:OOPSLA},
 Verifast \cite{Jacobs:POPL:2011}, and ShareInfer \cite{Le-Hobor:18}.
Unlike ours, these works support CSL with only one separating conjunction
and are much simpler than ours. 
%
Smallfoot discussed local reasoning for concurrency in the original CSL
\cite{OHearn:CON:2004}. As the logic reused the separating conjunction
to represent the ownership over separating resources, it could be
implied that the compositional analysis might be obtained straightforwardly
using the frame inference presented in \cite{Berdine:APLAS05}. Indeed,
such a frame inference was studied in jStar.
Furthermore, as shown in \cite{Brookes:CON:2004} and \cite{Bornat:POPL:2005},
the ownership transfer paradigm cannot
 describe {\em passivity} in programs that
 share read access among threads.
Verifast is a mechanised analyser
built on CSL with fractional permission \cite{Boyland:SAS:2003}.
ShareInfer, a bi-abductive procedure,
was based on the logic that
revamps the disjointness of permission heap regions
using tree shares. 
In contrast to ShareInfer, the underlying logic helps {\toolname} to
reason programs beyond precise  formulas. Furthermore,
it is a modular verification, which goes beyond
a frame inference procedure. 
While our system, like Smallfoot, jStar, and ShareInfer,
could be fully automated, we envision extending our frame inference with
abduction \cite{Calcagno:JACM:2011,Curry:ICECCS:2019,Le:CAV:2014,Trinh:APLAS:2013} and lazy initialization \cite{Pham:FM:2019,Pham:ATVA:2019} to infer procedure's 
 and a thread-modular summary. 
Additionally, we see potential in enhancing reasoning over inductive predicates
with an induction method, such as cyclic proofs like the one in \cite{Le:Frocas:2023,Loc:TACAS:2018}.
These future research directions demonstrate the scalability and adaptability of our system.

Our work, which develops an automatic verification
for concurrent programs, shares the same motivation with SmallfootRG \cite{SmallfootRG:SAS:2007}, Starling \cite{Windsor:CAV:2017}
and Caper \cite{Dinsdale:ESOP:2017}.
These systems rely on ordinary separation logic solvers
 (like GRASShopper
 \cite{Piskac:TACAS:2014}), it takes them a high cost to translate concurrent verification
 conditions. In contrast, {\toolname} builds a solver for $\slalp$
 so that it can discharge the verification conditions directly and is thus more efficient.

SmallfootRG is built on an extension of CSL with rely/guarantee and can verify
fine-grained concurrent programs. However, as it is based on the paradigm of
rely/guarantee, it maintains auxiliary global variables.
Hence, it could not preserve the local reasoning,
 the basis for compositional verification, like ours.
%
Starling is an instantiation of the Views framework \cite{Dinsdale:POPL:2013};
it computes for
each thread a separating view and then combines them, via a concurrency rule,
for the view of
their parent thread.
Starling neither supports functional correctness, frame inference, or modular verification.
Caper is based on concurrent abstract predicates \cite{Dinsdale:ECOOP:2010},
a separation logic with 
auxiliary guard algebras, shared regions, and actions. Although it can verify
the functional properties of
concurrent programs, unlike ours,
the paper did not present a proof system to infer frame assertions. Therefore,
it is unclear how Caper can compositionally reason about
programs.
 Finally, Diaframe implements a mechanised proof system based on Iris \cite{Mulder:PLDI:2022}.
 While its proof rules are based on bi-abduction concept, our system develop
 frame inference, the foundation for a bi-abductive inference for $\slalp$.

\paragraph{Conclusion} \label{sec.concl}
We develop foundational principles for the distribution between
strong and weak separating conjunctions.
Based on that, we build the frame inference procedure {\entname}
and the compositional verifier {\toolname}.
The proposed principles enable our system to verify fine-grained concurrent programs
which are beyond the current systems.
One of our future works is to extend {\entname}
with bi-abduction to infer specifications for procedures
and concurrent threads. Another direction is to explore cyclic proofs  \cite{Brotherston:MFPS:2025} for
the frame inference problem in $\slalp$ to enhance
the reasoning on looping programs.
We are also interested in testing the resulting
cyclic system with
real-world Move smart contracts \cite{Dill:CAV:2022}.
Yet another complementary direction is to study an under-approximation
of {\cslp} to find concurrency bugs.
These  works would advance
the verification of presence and absence of bugs in concurrent programs.


\bibliography{refs}


 \appendix

\section{Semantics}\label{csl-semantics}
The semantics is defined by structural induction on $\Phi$
as follows.
%
\[\begin{array}{l@{\hspace{0.25cm}}c@{\hspace{0.3cm}}l}
s,h,\rho \models \emp & \iff & \dom{h} = \emptyset \\
s,h,\rho \models x {\mapsto} y_1,..,y_n & \iff & \dom{h}=\{s(x)\} \mbox{ and } h(s(x))=(s(y_1),..,s(y_n)) \\
s,h,\rho \models P(\vec{x}) & \iff & (s(\vec{x}),h) \in \sem{P} \\
s,h,\rho \models A_1 * A_2 & \iff & \exists h_1, h_2.\ h = h_1 \circ h_2 \mbox{,} s,h_1,\rho \models A_1 \mbox{ and } s,h_2,\rho \models A_2 \\
s,h,\rho \models A_1 \wstar A_2 & \iff & \exists h_1, h_2.\ h = h_1 \wcirc h_2 \mbox{,} s,h_1,\rho \models A_1 \mbox{ and } s,h_2,\rho \models A_2 \\
s,h,\rho \models \heap \wedge \pure & \iff & s,h,\rho \models \heap  \mbox{ and } s,h,\rho \models \pure \\ 
s,h,\rho \models A^\pi & \iff &
\exists h'.\ h = \pi\cdot h' \mbox{ and } s,h',\rho \models A \\ 
s,h,\rho \models @_\alpha A & \iff & h = \rho(\alpha) \mbox{ and }  s,\rho(\alpha),\rho \models A \\
s,h,\rho \models \exists x.~ A & \iff &
\exists k.\ s[k \leftarrow x],h,\rho \models A \\
s,h,\rho \models \Phi_1 \lor \Phi_2 & \iff &
s,h,\rho \models \Phi_1 \mbox{ or } s,h,\rho \models \Phi_2 \\
s,h,\rho \models \lab_1=\lab_2 & \iff &
 \rho(\lab_1)=\rho(\lab_2) \\
s,h,\rho \models \lab_2 \disjoint \lab_3 & \iff &
\rho(\lab_2) \mbox{ and } \rho(\lab_3) \mbox{ are disjoint}  
\end{array}\]

\section{Proofs of \cref{lemma.sw.distribution}}
\subsection{Proof of \eqref{swdist}}
\begin{proof}
  \noindent  Assume $s,h, \rho \models (@_{\alpha} A \wstar @_{\beta} B) \sep (@_{\gamma} C \wstar @_{\xi} D)$, we have
  $h = (\rho(\alpha) \wcirc \rho(\beta)) \circ (\rho(\gamma) \wcirc \rho(\xi))$.
 This implies $\rho(\alpha) \wcirc \rho(\beta))$ and $\rho(\gamma) \wcirc \rho(\xi)$
  are disjoint. Consequently, we can infer
  \begin{align}
  s,h, \rho \models (\alpha \wcirc \beta) \bot (\gamma \wcirc \xi) \label{swdist.p1}
  \end{align}
We can also infer that $\rho(\alpha)$ and $\rho(\gamma)$ are disjoint, $\rho(\beta)$ and $\rho(\xi)$ are disjoint.
Thus, it implies $\rho(\alpha) \circ \rho(\gamma)$ and $\rho(\beta) \circ \rho(\xi)$
are defined.

For any ${\ell \in \dom{h}}$, we do the case split.
\begin{itemize}
\item Case left-1:  ${\ell \not \in \dom{\rho(\gamma) \wcirc \rho(\xi)}}$. This means
   ${\ell \not \in \dom{\rho(\gamma)}}$ and ${\ell \not \in \dom{ \rho(\xi)}}$.
 \begin{itemize}
  \item Case left-1.1:  ${\ell \not \in \dom{\rho(\beta)}}$, then $h(\ell) = \rho(\alpha)(\ell)$.
  \item Case left-1.2: ${\ell \not \in \dom{\rho(\alpha)}}$, then $h(\ell) = \rho(\beta)(\ell)$.
  \item Case left-1.3: $\rho(\alpha)(\ell) = (v, \pi)$ and $\rho(\beta)(\ell) = (v, \pi')$, then
     $h(\ell) = (v, \pi\otimes \pi')$.
 \end{itemize}
\item Case left-2:  ${\ell \not \in \dom{\rho(\alpha) \wcirc \rho(\beta)}}$. This means
   ${\ell \not \in \dom{\rho(\alpha)}}$ and ${\ell \not \in \dom{ \rho(\beta)}}$.
 \begin{itemize}
  \item Case left-2.1:  ${\ell \not \in \dom{\rho(\xi)}}$, then $h(\ell) = \rho(\gamma)(\ell)$.
  \item Case left-2.2: ${\ell \not \in \dom{\rho(\gamma)}}$, then $h(\ell) = \rho(\xi)(\ell)$.
  \item Case left-2.3: $\rho(\gamma)(\ell) = (v, \pi)$ and $\rho(\xi)(\ell) = (v, \pi')$, then
     $h(\ell) = (v, \pi\otimes \pi')$.
 \end{itemize}
\end{itemize}

Now, we re-arrange these cases into the following groups.
\begin{itemize}
\item Group right-1: includes Case left-1.1, and Case left-2.1 : ${\ell \not \in \dom{\rho(\beta)}}$ and  ${\ell \not \in \dom{\rho(\xi)}}$
  \begin{itemize}
   \item Group right-1.1: corresponds to Case left-1.1: ${\ell \not \in \dom{\rho(\gamma)}}$
   \item Group right-1.2: corresponds to Case left-2.1: ${\ell \not \in \dom{\rho(\alpha)}}$
   \end{itemize}
\item Group right-2: includes Case left-1.2, and Case left-2.2 : ${\ell \not \in \dom{\rho(\alpha)}}$ and  ${\ell \not \in \dom{\rho(\gamma)}}$
  \begin{itemize}
    \item Group right-2.1: corresponds to Case left-1.2: ${\ell \not \in \dom{\rho(\xi)}}$
    \item Group right-2.2: corresponds to Case left-2.2: ${\ell \not \in \dom{\rho(\beta)}}$
  \end{itemize}
  \item Group right-3: includes Case left-1.3, and Case left-2.3: $\ell$ in the overlapping region of $\dom{\rho(\alpha) \circ \rho(\gamma)}$ and $\dom{\rho(\beta) \circ \rho(\xi)}$
  \begin{itemize}
    \item Group right-3.1: $\ell$ in the overlapping region of $\dom{\rho(\alpha)}$ and $\dom{\rho(\beta)}$: this group corresponds to Case left-1.3.
    \item Group right-3.2: $\ell$ in the overlapping region of $\dom{\rho(\alpha)}$ and $\dom{\rho(\xi)}$: this group is empty as $\dom{\rho(\alpha)}$  and $\dom{\rho(\xi)}$ are disjoint.
    \item Group right-3.3: $\ell$ in the overlapping region of $\dom{\rho(\gamma)}$ and $\dom{\rho(\beta)}$: this group is empty as $\dom{\rho(\gamma)}$  and $\dom{\rho(\beta)}$ are disjoint.
    \item Group right-3.4: $\ell$ in the overlapping region of $\dom{\rho(\gamma)}$ and $\dom{\rho(\xi)}$:  this group corresponds to Case left-2.3.
  \end{itemize}
\end{itemize}
Group right-1, Group right-2 and Group right-3 show that
$h = (\rho(\alpha) \circ \rho(\gamma)) \wcirc (\rho(\beta) \circ \rho(\xi))$. This implies
  \begin{align}
  s,h, \rho \models (@_{\alpha} A \sep @_{\gamma} C) \wstar (@_{\beta} B \sep @_{\xi} D) \label{swdist.p2}
  \end{align}

  From \eqref{swdist.p1}, \eqref{swdist.p2} we have:
  $$s,h, \rho \models (@_{\alpha} A \sep @_{\gamma} C) \wstar (@_{\beta} B \sep @_{\xi} D)
    \land (\alpha \uweak \beta) \disjoint (\gamma \uweak \xi)$$
\end{proof}

\subsection{Proof of \eqref{wsdist}}

\begin{proof}
 \noindent  Assume $s,h, \rho \models (@_{\alpha} A \sep @_{\gamma} C) \wstar (@_{\beta} B \sep @_{\xi} D)
 \land (\alpha \uweak \beta) \disjoint (\gamma \uweak \xi)$,
 we have
 $h = (h_1 \circ h_3) \wcirc (h_2 \circ h_4)$
 with $h_1 = \rho(\alpha)$, $h_2 = \rho(\beta)$, $h_3 = \rho(\gamma)$ and $h_4 = \rho(\xi)$.

 As  $s,h, \rho \models  (\alpha \uweak \beta) \disjoint (\gamma \uweak \xi)$, we have
 $h_1 \wcirc h_2$ and $h_3 \wcirc h_4$
 are defined, and $h_1$ is disjoint from both $h_3$ and
 $h_4$, and $h_2$ is also disjoint from both $h_3$ and
 $h_4$.

 For any ${\ell \in \dom{(h_1 \circ h_3) \wcirc (h_2 \circ h_4)}}$, we do the case split.
 \begin{itemize}
\item Case left-1: ${\ell \not \in \dom{h_2 \circ h_4}}$. This means  ${\ell \not \in \dom{h_2}}$ and ${\ell \not \in \dom{h_4}}$.
  \begin{itemize}
   \item Case left-1.1:  ${\ell \not \in \dom{h_3}}$. Then, $h(\ell) = h_1(\ell)$.
   \item Case left-1.2: ${\ell \not \in \dom{h_1}}$. Then, $h(\ell) = h_3(\ell)$.
   \end{itemize}
\item Case left-2: ${\ell \not \in \dom{h_1 \circ h_3}}$. This means  ${\ell \not \in \dom{h_1}}$ and ${\ell \not \in \dom{h_3}}$.
  \begin{itemize}
    \item Case left-2.1: $\ell \not \in \dom{h_4}$. Then, $h(\ell) = h_2(\ell)$.
    \item Case left-2.2: $\ell \not \in \dom{h_2}$. Then, $h(\ell) = h_4(\ell)$.
  \end{itemize}
  \item Case left-3:  $\ell$ in the overlapping region of $\dom{h_1 \circ h_3}$ and $\dom{h_2 \circ h_4}$.
  \begin{itemize}
  \item Case left-3.1: $\ell$ in the overlapping region of $\dom{h_1}$ and $\dom{h_2}$.
     Assume $h_1(\ell) = (v, \pi)$ and $h_2(\ell) = (v, \pi')$, then
     $h(\ell) = (v, \pi\otimes \pi')$.
    \item Case left-3.2: $\ell$ in the overlapping region of $\dom{h_1}$ and $\dom{h_4}$. This case can not happen as $h_1$  and $h_4$ are disjoint.
    \item Case left-3.3: $\ell$ in the overlapping region of $\dom{h_3}$ and $\dom{h_2}$. This case can not happen as $h_3$  and $h_2$ are disjoint.
    \item Case left-3.4: $\ell$ in the overlapping region of $\dom{h_3}$ and $\dom{h_4}$. Assume $h_3(\ell) = (v, \pi)$ and $h_4(\ell) = (v, \pi')$, then
     $h(\ell) = (v, \pi\otimes \pi')$.
  \end{itemize}
 \end{itemize}

 Now, we re-arrange these cases into the following groups.
 \begin{itemize}
 \item Case right-1:   $\ell \not \in \dom{h_3 \wcirc h_4}$. This means $\ell \not \in \dom{h_3}$ and $\ell \not \in \dom{h_4}$. This includes Case left-1.1, Case left-2.1, and Case left-3.1.
   .
   \begin{itemize}
   \item Case right-1.1: $h(\ell) = h_1(\ell)$ where $\ell \not \in \dom{h_2}$ (this case corresponds to Case left-1.1).
      
   \item Case right-1.2: $h(\ell) = h_2(\ell)$ where $\ell \not \in \dom{h_1}$ (this case corresponds to Case left-2.1).
   \item Case right-1.3: $h(\ell) = (v, \pi\otimes \pi')$ where $h_1(\ell) = (v, \pi)$ and $h_2(\ell) = (v, \pi')$ (this case corresponds to Case left-3.1).
 \end{itemize}   
 \item Case right-2:
     $\ell \not \in \dom{h_1 \wcirc h_2}$. This means $\ell \not \in \dom{h_1}$ and $\ell \not \in \dom{h_2}$. This includes Case left-1.2, Case left-2.2, and Case left-3.4.
   .
   \begin{itemize}
   \item Case right-2.1: $h(\ell) = h_3(\ell)$ where $\ell \not \in \dom{h_4}$ (this case corresponds to Case left-1.2).
      
   \item Case right-2.2: $h(\ell) = h_4(\ell)$ where $\ell \not \in \dom{h_3}$ (this case corresponds to Case left-2.2).
   \item Case right-2.3: $h(\ell) = (v, \pi\otimes \pi')$ where $h_3(\ell) = (v, \pi)$ and $h_4(\ell) = (v, \pi')$ (this case corresponds to Case left-3.4).
 \end{itemize}   
 \end{itemize}

 Based on the re-arranged cases, we have $h = (h_1 \wcirc h_2) \circ (h_3 \wcirc h_4)$.
 Thus,  $s,h, \rho \models (@_{\alpha} A \wstar @_{\beta} B) \sep (@_{\gamma} C \wstar @_{\xi} D)$.
\end{proof}

\section{Proofs of \cref{splitjoin}}
\subsection{Proof of \eqref{wsplit}}
\begin{proof}
  We prove
   \begin{align}
    (@_\alpha A)^{\pi_1 \oplus \pi_2}  & \models (@_\alpha A)^{\pi_1} \wstar
    (@_\alpha A)^{\pi_2}
     \tag{$\wstar$ Split}
   \end{align}

   Suppose that $s,h, \rho \models (@_\alpha A)^{\pi_1 \oplus \pi_2}$.
   We have $h = (\pi_1\oplus \pi_2) \cdot h'$
   where $s,h', \rho \models (@_\alpha A)$ (1).
   By semantics of the heap label, we have
   $h' = \rho(\alpha)$ and $s,h', \rho \models A$.
   And for any $\ell \in \dom{h}$, we have $h'(\ell) = (v, \pi)$ (2).
   
   (1) and (2) imply $h(\ell) = (v, (\pi_1\oplus \pi_2) \otimes \pi)$.
   
   And using the permission algebra axiom on distribution in \cref{sec:language} ($\forall \pi, \pi_1, \pi_2 \in \perm.~(\pi_1\oplus\pi_2)\otimes \pi = (\pi_1\otimes\pi)\oplus (\pi_1\oplus\pi)$), we obtain
   $$h(\ell) = (v, (\pi_1 \otimes \pi) \oplus (\pi_2 \otimes \pi))$$
   Now we define $h_1$ and $h_2$ where $\dom{h_1} = \dom{h_2} = \dom{h}$ and
   $h_1(\ell) = (v, (\pi_1 \otimes \pi))$ and $h_2(\ell) = (v, (\pi_2 \otimes \pi))$.
   Hence,  $h_1 = \pi_1 \cdot h'$ (3a),  $h_2 = \pi_2 \cdot h'$ (3b), $h = h_1 \wcirc h_2$ (3c).

   From (1) and (3a), we have $s,h_1, \rho \models (@_\alpha A)^ {\pi_1}$.

   From (1) and (3b), we have $s,h_2, \rho \models (@_\alpha A)^ {\pi_2}$.

   Combined with (3c) and from the semantics of weak separation, we have
   $s,h, \rho \models  (@_\alpha A)^ {\pi_1} \wstar  (@_\alpha A)^ {\pi_2}$.   
\end{proof}

\subsection{Proof of \eqref{wjoin}}
\begin{proof}
  We prove
  \begin{align}
      (@_\alpha A)^{\pi_1} \wstar
 (@_\alpha A)^{\pi_2} & \models (@_\alpha A)^{\pi_1 \oplus \pi_2}
  \tag{$\wstar$ Join}
  \end{align}
  Suppose that $s,h, \rho \models (@_\alpha A)^{\pi_1} \wstar (@_\alpha A)^{\pi_2}$.
  By the semantics of weak separation, there exists $h_1$, $h_2$,
  $h = h_1 \wcirc h_2$ such that $s,h_1, \rho \models (@_\alpha A)^ {\pi_1}$
  and $s,h_2, \rho \models (@_\alpha A)^ {\pi_2}$.

  Furthermore, we have $h_1 = \pi_1 \cdot h_1'$
  where $s,h_1', \rho \models (@_\alpha A)$ and $h_1'=\rho(\alpha)$(1).
  Similarly, we have $h_2 = \pi_2 \cdot h_2'$
  where $s,h_2', \rho \models (@_\alpha A)$ and $h_2'=\rho(\alpha)$(2).

  From (1) and (2), we have $h_1' = h_2' = \rho(\alpha)$.
  This gives us $h = (\pi_1 \cdot h_1') \wcirc (\pi_2 \cdot h_1')$ .

  For any $\ell \in \dom{h_1'}$ i.e., $h_1'(\ell) = (v, \pi)$.
  Now we define $h_a$ and $h_b$ where $\dom{h_a} = \dom{h_b} = \dom{h_1'}$ and
  $h_a(\ell) = (v, (\pi_1 \otimes \pi))$ and $h_b(\ell) = (v, (\pi_2 \otimes \pi))$.
  This gives us $((\pi_1 \cdot h_1') \wcirc (\pi_2 \cdot h_1'))(\ell) = (h_a \wcirc h_b)(\ell)
  = (v, (\pi_1 \otimes \pi) \oplus (\pi_2 \otimes \pi) )$ = $(v, (\pi_1 \oplus \pi_2) \otimes \pi)$ = $((\pi_1 \oplus \pi_2) \cdot h_1')(l) $ (3).

  (1) and (3) give us $s,h, \rho \models (@_\alpha A)^{\pi_1 \oplus \pi_2}$.
\end{proof}

\section{Proofs of {\inwardtext} and {\inwardstartext}}
\subsection{Proof of \eqref{label.perm}}
\begin{proof}
 We prove
  \begin{align}
(@_\alpha A)^{\pi}  & \equiv  @_{ \alpha^\pi} A^{\pi} \tag{$@^{\pi}$}
  \end{align}

  Suppose that $s,h, \rho \models (@_\alpha A)^{\pi}$.
  The semantics of permission regions gives us: $h = \pi \cdot h'$
  and $s,h', \rho \models @_\alpha A $.
  Furthermore, the semantics of labels gives us: $h' = \rho(\alpha)$
  and  $s,h', \rho \models A $.
  Then  $h = \pi \cdot \rho(\alpha) = \rho(\alpha^\pi)$
  and $s,h, \rho \models A^\pi $.
  This gives us $s,h, \rho \models @_{ \alpha^\pi} A^{\pi}$, as requested.
  \end{proof}
\subsection{Proof of \eqref{sep.pi}}
\begin{proof}
  First, we prove
  \begin{align}
(A \sep B)^{\pi}  & \equiv  A^{\pi} \sep B^{\pi} \tag{a} \label{a}
  \end{align}

   Suppose that $s,h, \rho \models (A \sep  B)^{\pi}$.
  We have $h = \pi \cdot h_1$
  and $s,h_1, \rho \models A \sep  B$.
  By definition of strong separation,
  $h_1 = h_a \circ h_b$
  and $s,h_a, \rho \models A$, $s,h_b, \rho \models B$.
  This gives us $s, \pi \cdot h_a, \rho \models A^\pi$, $s, \pi \cdot h_b, \rho \models B^\pi$
  (1).

$h = \pi \cdot h_1 = \pi \cdot (h_a \circ h_b)$
  
  For any $\ell \in \dom{h_1}$ i.e., $h_1(\ell) = (v, \pi')$,
  we have $h(\ell) = (v, \pi \otimes \pi')$.

  We consider two cases.
  \begin{enumerate}
  \item Case 1: $\ell \in \dom{h_a}$ and $\ell \not\in \dom{h_b}$.
    Suppose $h_1 (\ell) = h_a(\ell) = (v_a, \pi_1)$.
    Then,  $h (\ell) =  (v_a, \pi_1 \otimes \pi) = (\pi \cdot h_a)(\ell)$.
    Since  $\ell \not\in \dom{h_b}$, $h (\ell) = ((\pi \cdot h_a)\circ (\pi \cdot h_b) )(\ell)$(3).

    (1), (3), and the semantics of strong separation give us
    $s,h, \rho \models A^{\pi} \sep B^{\pi}$.
  \item Case 2: $\ell \in \dom{h_b}$ and $\ell \not\in \dom{h_a}$.
    This case is symmetry to Case 1.
    \end{enumerate}

  Next, we prove
  \begin{align}
@_\gamma (@_\alpha A \sep @_{\beta} B)  \wedge \gamma = \alpha \udisjoint \beta &\equiv @_{\alpha} A \sep @_{\beta} B  \wedge \gamma = \alpha \udisjoint \beta \tag{b} \label{b}
  \end{align}

  The proof of $@_\gamma (@_{\alpha} A \sep @_{\beta} B)  \wedge \gamma = \alpha \udisjoint \beta \models @_{\alpha} A \sep @_{\beta} B  \wedge \gamma = \alpha \udisjoint \beta$ is easy.
  We show the proof of $ @_{\alpha} A \sep @_{\beta} B  \wedge \gamma = \alpha \udisjoint \beta \models @_\gamma (@_{\alpha} A \sep @_{\beta} B)  \wedge \gamma = \alpha \udisjoint \beta$.

  Suppose that $s, h , \rho \models @_{\alpha} A \sep @_{\beta} B  \wedge \gamma = \alpha \udisjoint \beta$.
  By the semantics of conjunction,
   $s, h , \rho \models @_{\alpha} A \sep @_{\beta} B$ (4a) and $ s, h , \rho \models \gamma = \alpha \udisjoint \beta$ (5).  Then by strong separating conjunction, $h = h_1 \udisjoint h_2$
  and $s, h_1 , \rho \models @_{\alpha} A$ and $s, h_2 , \rho \models @_{\beta} B$.
  Then $h_1 = \rho(\alpha)$, $h_2 = \rho(\beta)$. Hence $h = h_1 \udisjoint h_2 = \rho(\alpha) \udisjoint \rho(\beta) = \rho(\alpha \udisjoint \beta) = \rho(\gamma)$ (6).

  (4a) and (6) give us $s, h , \rho \models @_\gamma (@_{\alpha} A \sep @_{\beta} B)$ (4b) (by the semantics of label).
  This results and (5) give us $s, h , \rho \models @_\gamma (@_{\alpha} A \sep @_{\beta} B) \wedge \gamma = \alpha \udisjoint \beta $ (by the semantics of the conjunction).

  Now, for the main result, let $s,h$ and $\rho$ be given. We have
  \[
  \begin{array}{clr}
    & s, h, \rho \models @_\gamma (@_\alpha A \sep @_\beta B)^{\pi}  \wedge \gamma = \alpha^\pi \udisjoint \beta^\pi & \\
    \Leftrightarrow & s, h, \rho \models @_\gamma ((@_\alpha A)^\pi \sep (@_\beta B)^\pi )  \wedge \gamma = \alpha^\pi \udisjoint \beta^\pi& (\mbox{by (\ref{a})}) \\
    \Leftrightarrow & s, h, \rho \models @_\gamma (@_{\alpha^\pi} A^\pi \sep @_{\beta^\pi} B^\pi)  \wedge \gamma = \alpha^\pi \udisjoint \beta^\pi& (\mbox{by (\ref{label.perm}) and } \sep)\\
    \Leftrightarrow & s, h, \rho \models @_{\alpha^\pi} A^\pi \sep @_{\beta^\pi} B^\pi  \wedge \gamma = \alpha^\pi \udisjoint \beta^\pi& (\mbox{by (\ref{b})}) 
  \end{array}
  \]
\end{proof}

\subsection{Proof of \eqref{wsep.pi}}
\begin{proof}
   First, we prove
  \begin{align}
    (A \wstar  B)^{\pi}  & \models  A^{\pi} \wstar B^{\pi} \tag{d} \label{d}
    \end{align}
\end{proof}
  Suppose that $s,h, \rho \models (A \wstar  B)^{\pi}$.
  We have $h = \pi \cdot h_1$
  and $s,h_1, \rho \models A \wstar  B$.
  By definition of weak separation,
  $h_1 = h_a \wcirc h_b$
  and $s,h_a, \rho \models A$, $s,h_b, \rho \models B$.
  This gives us $s, \pi \cdot h_a, \rho \models A^\pi$, $s, \pi \cdot h_b, \rho \models B^\pi$
  (1).

$h = \pi \cdot h_1 = \pi \cdot (h_a \wcirc h_b)$
  
  For any $\ell \in \dom{h_1}$ i.e., $h_1(\ell) = (v, \pi')$,
  we have $h(\ell) = (v, \pi \otimes \pi')$.

  We consider three cases.
  \begin{enumerate}
  \item Case 1: $\ell \in \dom{h_a} \cap \dom{h_b}$ and $h_a(\ell) = (v, \pi_1)$,
    $h_b(\ell) = (v, \pi_2)$,  $h_1(\ell) = (v, \pi_1 \oplus \pi_2)$.
    This gives us $h(\ell) = (v, (\pi_1 \oplus \pi_2) \otimes \pi)$.

    And using the permission algebra axiom on distribution in \cref{sec:language},
    we get $h(\ell) = (v, (\pi_1 \otimes \pi) \oplus (\pi_2 \otimes \pi) )$.
    This implies
    $h = (\pi \cdot h_a) \wcirc (\pi \cdot h_b) $ (2).

    (1), (2), and the semantics of weak separation give us
    $s,h, \rho \models A^{\pi} \wstar B^{\pi}$.
  \item Case 2: $\ell \in \dom{h_a}$ and $\ell \not\in \dom{h_b}$.
    Suppose $h_1 (\ell) = h_a(\ell) = (v_a, \pi_1)$.
    Then,  $h (\ell) =  (v_a, \pi_1 \otimes \pi) = (\pi \cdot h_a)(\ell)$.
    Since  $\ell \not\in \dom{h_b}$, $h (\ell) = ((\pi \cdot h_a)\wcirc (\pi \cdot h_b) )(\ell)$(3).

    (1), (3), and the semantics of weak separation give us
    $s,h, \rho \models A^{\pi} \wstar B^{\pi}$.
  \item Case 3: $\ell \in \dom{h_b}$ and $\ell \not\in \dom{h_a}$.
    This case is symmetry to Case 2.
  \end{enumerate}

  Now, for the main result, let $s,h$ and $\rho$ be given. We have
  \[
  \begin{array}{clr}
    & s, h, \rho \models @_\gamma (@_\alpha A \wstar @_\beta B)^{\pi}  \wedge \gamma = \alpha^\pi \uweak \beta^\pi & \\
    \Rightarrow & s, h, \rho \models @_\gamma ((@_\alpha A)^\pi \wstar (@_\beta B)^\pi )  \wedge \gamma = \alpha^\pi \uweak \beta^\pi& (\mbox{by (\ref{d})}) \\
    \Leftrightarrow & s, h, \rho \models @_\gamma (@_{\alpha^\pi} A^\pi \wstar @_{\beta^\pi} B^\pi)  \wedge \gamma = \alpha^\pi \uweak \beta^\pi& (\mbox{by (\ref{label.perm}) and } \wstar)\\
    \Leftrightarrow & s, h, \rho \models @_{\alpha^\pi} A^\pi \wstar @_{\beta^\pi} B^\pi  \wedge \gamma = \alpha^\pi \uweak \beta^\pi& (\mbox{by weakening}) 
  \end{array}
  \]
\subsection{Proof of \eqref{pdown}}

\begin{proof}
  We prove
  \begin{align}
    (A^{\pi_1})^{\pi_2}  & \models A^{\pi_1 \otimes \pi_2} &
  \tag{$\otimes$}
  \end{align}

  Suppose that $s,h, \rho \models (A^{\pi_1})^{\pi_2}$.
  We have $h_1 = \pi_2 \cdot h$
  where $s,h_1, \rho \models A^{\pi_1}$.
  Again, We have $h_2 = \pi_1 \cdot h_1 = \pi_1 \cdot \pi_2 \cdot h$
  where $s,h_2, \rho \models A$ (1).

  For any $\ell \in \dom{h}$ i.e., $h(\ell) = (v, \pi)$,
  by definition of the multiplication, we have $(\pi_1 \cdot \pi_2 \cdot h) = (v, \pi \otimes \pi_1 \otimes \pi_2)$ (2).

  (1) and (2) give us $s,h, \rho \models A ^{\pi_1 \otimes \pi_2}$.
\end{proof}

\end{document}